\documentclass[twocolumn]{openjournal}

\usepackage{amsmath}
\usepackage{amssymb}
\usepackage{graphicx} 
\usepackage[dvipsnames]{xcolor}
\usepackage{hyperref}
\usepackage{multirow}
\usepackage{soul}
\usepackage{floatrow}
\usepackage{blindtext}

\defcitealias{Mandelbaum2018}{SRD}

\newcommand{\kmax}{k_{\rm max}}
\newcommand{\bs}{b_{\rm s^2}}
\newcommand{\blap}{b_{\nabla^2}}
\newcommand{\logMc}{\text{log}_{10}M_{\rm C}}
\newfloatcommand{capbtabbox}{table}[][\FBwidth]

\usepackage[english]{babel}
\addto\extrasenglish{%
  
}

\newcommand{\mgl}{\texttt{MGL}} 
\newcommand{\bacco}{\texttt{BACCO}}

\begin{document}
\journalinfo{The Open Journal of Astrophysics}

\title{Balancing bias, baryons, and scale cuts in LSST $3\times2$pt analysis}
\email[$^\star$: Corresponding author: ]{ottavia.truttero@ed.ac.uk}

\author{Ottavia~Truttero$^{\star a}$}
\author{Maria Tsedrik$^{a,b}$}
\author{Joe~Zuntz$^{a}$}
\author{Alkistis~Pourtsidou$^{a,b}$}
\author{Nikolina \v{S}ar\v{c}evi\'c$^{c}$}
\author{Christos~Georgiou$^{d}$}
\author{the LSST Dark Energy Science Collaboration}

\affiliation{$^{a}$ Institute for Astronomy, University of Edinburgh, Royal Observatory, Blackford Hill, Edinburgh, EH9 3HJ, U.K.}
\affiliation{$^{b}$Higgs Centre for Theoretical Physics, School of Physics and Astronomy, University of Edinburgh, Edinburgh EH9 3FD, U.K.}
\affiliation{$^{c}$Department of Physics, Duke University, Durham, NC 27708, USA}
\affiliation{$^{d}$ Institut de F\'{i}sica d'Altes Energies (IFAE), The Barcelona Institute of Science and Technology, Campus UAB, 08193 Bellaterra, Barcelona, Spain}

\begin{abstract}
    Stage IV surveys such as LSST will probe deeply into the nonlinear regime, where systematic effects from galaxy bias and baryonic feedback become dominant and poorly constrained nuisance parameters can lead to degeneracies. 
    In this work we present a $3\times2$pt analysis for LSST Y1 and Y10 data using the \bacco ~emulator for modelling both the hybrid-effective field theory (HEFT) for nonlinear galaxy bias and the baryonic feedback using the baryonification mechanism.
    We aim to find a balance between model complexity and scale cuts, with particular attention to parameter degeneracies and baryonic feedback effects on the galaxy--matter and galaxy--galaxy power spectra.  First, we find that a linear bias model delivers percent-level, unbiased constraints on $\Omega_{\rm m}$ and $\sigma_8$ only up to $\kmax=0.1\,h/$Mpc,
    but pushing to smaller scales requires a perturbative approach. Second, we compare HEFT with a minimal bias variant with fixed higher-order terms, and find that the latter is unbiased in $\Lambda$CDM even at $\kmax=0.7\,h/$Mpc.
    We show that higher-order bias can mimic baryonic suppression, but baryons cannot reproduce the full range of higher-order bias behaviour within the parameter range allowed by the \bacco ~baryonification model.
    Third, we find that a detection of the total neutrino mass $M_\nu$ is possible for both Y1 and Y10 for $k\geq0.5\,h/$Mpc, at least when photo-$z$ uncertainties and related nuisance parameters are precisely known. However, the specific measured value is not robust across equally plausible mock scenarios: the inferred $M_\nu$ can be significantly biased by adopting the minimal bias model.
    The entire analysis is conducted with a new independent, open source pipeline (\texttt{MGL}) that we present for the first time in this work.
\end{abstract}

\maketitle

\section{Introduction}
With the new cosmological surveys, we are witnessing the beginning of a new era of cosmology. Building on the foundations laid by Stage III surveys -- such as the Dark Energy survey (DES) \citep{Abbott2022}, the Kilo Degree survey (KiDS) \citep{Wright2024}, and the Hyper Suprime-Cam (HSC) \citep{Aihara2022} -- Stage IV photometric surveys -- including \textit{Euclid} \citep{EuclidCollaboration2025} and the Rubin Observatory Legacy Survey of Space and Time (LSST) \citep{Mandelbaum2018} -- will gain unprecedented statistical power and achieve significantly tighter constraints on dark matter and dark energy.
In addition to weak gravitational lensing, these surveys will also probe galaxy clustering over vast volumes and to high redshift, enabling tighter constraints on neutrino mass and other extensions or modifications to the standard $\Lambda$CDM cosmological model.

A key observable for these cosmological surveys is the so-called $3\times2$pt statistic, which refers to the joint analysis of three two-point correlation functions: the autocorrelation of galaxy positions, the autocorrelation of galaxy shapes, and their cross-correlation. Both galaxy positions (clustering) and shapes (gravitational weak lensing or shear) are tracers of the underlying dark matter distribution. As these two probes are sensitive to slightly different physics and different systematics, we benefit by their combination as it breaks parameter degeneracies, self-calibrates against systematic effects such as galaxy bias, intrinsic alignments, and photometric redshift errors, while boosting the overall signal-to-noise of cosmological inferences (see \textit{e.g.}, \citet{Krause2021, Heymans2021, Sarcevic2025, Zhang2025}).

Unlocking the full statistical power of LSST, as well as the other Stage IV surveys, requires more understanding of the small-scales systematics. On one hand, at scales smaller than $k\sim0.1\,h/$Mpc the redistribution of dark matter due to baryonic physics (\textit{i.e.}, the so-called baryonic feedback) becomes non negligible. The baryonic feedback effect propagates into both cosmic shear and galaxy clustering analysis, and it cannot be ignored without affecting the result on cosmological parameters \citep{Huang2018, Chisari2019, Schneider2020a, Bera2026}. Because the underlying physics is complex and not well understood, purely analytic descriptions are inadequate. Instead, mitigation relies on hydrodynamical simulations, \textit{e.g.}, the \texttt{FLAMINGO} Simulation \citep{Schaye2023}, and semi-empirical approaches such as Baryonic Correction Models or baryonification \citep{Schneider2015, Schneider2019, Arico2020}, and emulator-based models such as the \bacco ~emulator \citep{Arico2021}.

On the other hand, the way galaxies trace the underlying dark matter distribution (\textit{i.e.}, galaxy bias) also becomes nonlinear at nonlinear scales. 
In the previous experiments \citep{Abbott2018, Prat2022, Sugiyama2023} stringent scale cuts and linear bias prescription were applied to the photometric galaxy clustering and cross-correlation with shear, hence, modelling of the baryonic suppression was not required at these scales. However, recent developments in galaxy bias models, as effective field theory and the hybrid-effective field theory, have allowed to model bias down to scales of $k = 0.7\,h/$Mpc, where baryonic feedback also becomes not negligible \citep{Modi2020, Hadzhiyska2021, Nicola2024}. 

Therefore, in addition to the usual 6 cosmological parameters, in a $3\times2$pt analysis we must account for both modelling of nonlinear bias and baryonic feedback, which introduce two sets of nuisance parameters. These parameters are typically weakly constrained and can exhibit some degeneracies. In this work we study how baryonic and bias parameters affect the LSST $3\times2$pt data, with particular interest in potential degeneracies and their imprint on the power spectra. We present forecasts for LSST Year 1 (Y1) and Year 10 (Y10).

This work is organised as follows: the galaxy bias and baryonic feedback models employed in this work are described in \autoref{Sec:Systematics}, \autoref{Sec:Analysis} describes the methodology used in our analysis, including a presentation of a new pipeline (\mgl). We finally discuss our results in \autoref{Sec:Results} and conclude in \autoref{Sec:Conclusions}.

\section{Modelling systematics} \label{Sec:Systematics}
\subsection{Nonlinear bias} \label{Subsec:nl-bias}
Galaxies act as biased tracers of the underlying dark matter distribution \citep{Fry1993, McDonald2009, Desjacques2018}. At first approximation, the relation between galaxy overdensities, $\delta_{\rm g}$, and matter overdensities, $\delta_{\rm m}$, can be described using a simple linear bias model,
\begin{equation}
    \delta_{\rm g} = b^{\rm E}_1 \delta_{\rm m} \, ,
    \label{eq:linear_biasE}
\end{equation}
where $b^{\rm E}_1$ is the linear bias parameter in the Eulerian framework (E). By ``Eulerian'' we mean that all field evolutions are evaluated in a specific position and redshift, so bias parameters relate the galaxy field to the evolved matter field at the same location.

In Fourier space ($k$), the matter power spectrum $P_{\rm mm}(\textbf{k})$ is defined by the two-point correlation function
\begin{equation}
    \langle \delta^{*}_{\rm m}(\textbf{k}) \delta_{\rm m}(\textbf{k}') \rangle = (2\pi)^3\delta_D(\textbf{k}-\textbf{k}') P_{\rm mm}(k)\,,
\end{equation}
where $\delta_D$ indicates the Dirac delta function.
Thus, in the scenario described by \autoref{eq:linear_biasE}, the galaxy--galaxy power spectrum and galaxy--matter power spectrum relate to the matter power spectrum as
\begin{align}
    \begin{split}
        P_{\rm gg}(k,z) &= (b^{\rm E}_1)^2 P_{\rm mm}(k,z) \, ,\\
        P_{\rm gm}(k,z) &= b^{\rm E}_1 P_{\rm mm}(k,z) \, .
    \end{split}
\end{align}

The validity of these linear relations corresponds to the validity of the first order expansion in the matter density perturbations, \textit{i.e.}, up to $k\sim 0.1\,h/$Mpc. In a more general, nonlinear bias description, the connection between the overdensity fields can be described using perturbation theory, which allows for inclusion of quasi-nonlinear scales. For instance, in the Lagrangian approach, which relates the bias parameters to the initial density field as formulated in \citet{Matsubara2008}, the galaxy to matter correction is given by:
\begin{equation} 
    1 + \delta_{\rm g}(\textbf{x}) = \int d^3\textbf{q} \, w(\textbf{q}) \, \delta_{\rm D}(\textbf{x} - \textbf{q} - \Psi(\textbf{q})), 
    \label{eq:delta_g_integral} 
\end{equation}
where $\textbf{x}$ and $\textbf{q}$ denote the Eulerian and Lagrangian coordinates, respectively, while $\Psi(\textbf{q})$ is the Lagrangian displacement field. The weighting function $w$ in \autoref{eq:delta_g_integral} is a functional of linear combinations of the second derivatives of the gravitational potential $\Phi$.  
Specifically, there are four terms that contribute to the expansion at one-loop correction: the matter overdensity $\delta_{\rm m}$, which is proportional to the trace $\nabla^2\Phi$; the square of the matter overdensity, $\delta_{\rm m}^2$; the squared tidal field, $s^2 \equiv s_{ij}s^{ij}$, where $s_{ij} \equiv \partial_i\partial_j\Phi - \delta_{ij} \nabla^2\Phi/3$; and the Laplacian of the matter overdensity, $\nabla^2\delta_{\rm m}$, which contains the non-local aspects of galaxy formation \cite{Chan2012, Baldauf2012, Baldauf2013, Lazeyras2016}.
\autoref{eq:delta_g_integral} can be therefore expressed as a superposition of these fields weighted by their corresponding bias parameters,
\begin{align}
\begin{split}
    1 + \delta_{\rm g} &= 1 + b^{\rm L}_1\delta_{\rm m} + b^{\rm L}_2(\delta_{\rm m}^2 - \langle\delta_{\rm m}^2\rangle) \\&+ b^{\rm L}_{s^2}(s^2 - \langle s^2\rangle) + b^{\rm L}_{\nabla^2}\nabla^2\delta_{\rm m} + \epsilon,
    \end{split}
    \label{eq:galaxy_bias_expansion}
\end{align} 
where $\epsilon$ is a stochastic contribution that incorporates random initial conditions for galaxy formation, and the superscript ``L'' denotes Lagrangian framework.
The Eulerian and Lagrangian expansions are equivalent; there is a relation between their bias parameters, since galaxies in an evolved Eulerian field can be traced back to proto-galaxies in Lagrangian space at early times \citep{Matsubara2011}.

Following from \autoref{eq:galaxy_bias_expansion}, the galaxy--galaxy and galaxy--matter power spectra can be written as
\begin{align}
    \begin{split}
        P_{\rm gg}(k,z) &= \sum_{\alpha,\beta} b_\alpha b_\beta P_{\alpha\beta}(k,z) + P_{\rm SN}, \\
        P_{\rm gm}(k,z) &= \sum_\alpha b_\alpha P_{\alpha 1}(k,z),
    \end{split}
    \label{eq:Pgg_bias_expansion_general}
\end{align}
where $\alpha, \beta \in \{1, \delta_{\rm m}, \delta_{\rm m}^2, s^2, \nabla^2\delta_{\rm m}\}$, $b_1=1, b_{\delta_{\rm m}}=b_1^{\rm L}, b_{\delta_{\rm m}^2}=b_2^{\rm L},b_{s^2}=b_{s^2}^{\rm L}, b_{\nabla^2\delta_{\rm m}}=b_{\nabla^2}^{\rm L}$  and $P_{\rm SN}$ is the power spectrum of the stochastic term $\epsilon$ in the previous bias equation, which is assumed to be scale-independent. The expansion in \autoref{eq:Pgg_bias_expansion_general} results in 15 cross-spectra $P_{\alpha\beta}$ (one for each combination of $\alpha$ and $\beta$), which can be obtained either from perturbation theory \citep{McEwen2016, Chen2020} or hybrid approaches that combine simulations and perturbation theory \citep{Modi2020}. In this work we adopt the latter, using the \bacco ~emulator \citep{Arico2021a, Arico2021, Angulo2021, Arico2023, Zennaro2023, Pellejero2023, Burger2025, Pellejero2026} calibrated on \textit{N}-body simulations as described in \citet{Zennaro2023}.

While linear theory holds for scales $k\lesssim0.1\,h/$Mpc, the standard perturbative approach allows us to model scales $k\lesssim 0.2\,h/$Mpc \citep{Bernardeau2002}. 
The Effective Field Theory of Large-Scale Structure (EFTofLSS) \citep{Baumann2012, Carrasco2012, Pajer2013}, further extends this range by incorporating the effect of complex small-scales physics in mildly nonlinear effective terms. But, despite being an exact approach, it only allows a mild gain in scales ($k \lesssim0.4\,h/$Mpc in real space).
However, Stage IV surveys such as LSST \citep{Mandelbaum2018} will provide high-precision data even at smaller scales, where baryonic feedback becomes a dominant source of uncertainty. To take full advantage of this information, we require a more sophisticated approach. 

To this end, we adopt here the Hybrid Effective Field Theory (HEFT) approach \citep{Modi2020, Kokron2021, Hadzhiyska2021} with \bacco ~emulator. Here, the hybrid bias expansion consists of considering the relation between the galaxy overdensity and the gravitational potential perturbatively at early times, \autoref{eq:delta_g_integral}, and perform the subsequent evolution under gravity (\textit{i.e.}, solving for the Lagrangian displacement $\Psi$) via numerical $N$-body simulations. This approach significantly increases the accessible range of scales, $k\leq 0.7\,h/$Mpc.

\subsection{Baryonic feedback} \label{Subsec:baryonic-feedback}
To achieve an accurate interpretation of cosmological data in Stage IV surveys, baryonic feedback must be taken into account when using scales smaller than $\kmax\simeq0.2\,h/$Mpc \citep{Truttero2025}. It has been shown that ignoring these effects can lead to large biases in cosmological parameter estimation - particularly in $\Omega_{\rm m}$ and $\sigma_8$ \citep{Schneider2020a} - and may even lead to false detections of exotic physics \citep{Schneider2020b, Tsedrik2024}.

Among the available methods \citep{Huang2018}, we choose to use the \bacco ~emulator which includes baryonic effects in the nonlinear dark matter-only (DMO) power spectrum, $P_{\rm mm, DMO}$, through a baryonification algorithm. With this method, baryons are included in DMO $N$-body simulations by perturbatively shifting the particles in order to consider effects from gas, stars and active galactic nuclei. The result is a ``baryonified'' density profile, which is then used in the simulation to study the evolution of structures. 

\bacco ~is a neural network-based emulator which includes 7 additional baryonic parameters: the extent of the ejected gas ($\eta$), the density profiles of hot gas in haloes ($\theta_{\rm inn},\, M_{\rm inn},\, \theta_{\rm out}$), the fraction of gas retained in haloes of a given mass ($M_{\rm c},\, \beta$), and the characteristic halo mass scale for central galaxies ($M_{\rm 1,z0,cen}$). 
Across a wide range of cosmological hydrodynamical simulations, \bacco ~achieves an accuracy of $\sim 1$–$5\%$ \citep{Arico2021}. It covers scales from $0.01\,h/$Mpc to $5\,h/$Mpc, with constant extrapolation at smaller scales, and assumes a $\Lambda$CDM framework extended to include massive neutrinos and dynamical dark energy. 

While it has been shown that hydrodynamical simulations and Stage IV setups prefer time-dependent baryonic feedback, especially $\logMc$ \citep{Schneider2025, Giri2021, Wayland2026}, in this work all baryonic parameters are considered as redshift independent, following the \bacco ~model without time dependence.

The net impact of baryonic feedback is typically quantified by the ratio between the matter power spectrum when including baryonic physics, $P_{\rm mm,DMB}$, and that from a dark matter only one, $P_{\rm mm, DMO}$, \textit{i.e.}, by a suppression factor:
\begin{equation}
    S_{\rm mm}(k, z) = \frac{P_{\rm mm,DMB}(k, z)}{P_{\rm mm, DMO}(k, z)} \, .
\end{equation}
This factor is then applied as a correction to the nonlinear matter power spectrum prediction.

\citet{Zennaro2024} proposed an approximation calibrated on hydrodynamical simulations to include baryonic feedback in the galaxy-galaxy and the galaxy-matter power spectra within the perturbative framework:
\begin{align}
    \begin{split}
        P_{\rm gg}(k) &= P_{\rm gg,DMO}(k), \\
        P_{\rm gm}(k) &= \sqrt{S_{\rm mm}(k)} \, P_{\rm gm,DMO}(k),
    \end{split}
    \label{eq:Pgg_Pgm_zennaro}
\end{align}
\citet{Zennaro2024} tested the second relation by comparing $P_{\rm gm}$ against the FLAMINGO simulation, and found it to be accurate at the percent level for scales $k\lesssim0.7\,h/$Mpc, while the galaxy-galaxy power spectrum $P_{\rm gg}$ contains enough freedom in the nonlinear bias expansion to include the baryonic information at these scales. Both $P_{\rm gg,DMO}$ and $P_{\rm gm,DMO}$ are computed with \autoref{eq:Pgg_bias_expansion_general}. Hereafter we refer to this approach as \textit{Zennaro24}.

\section{Modelling and analysis} \label{Sec:Analysis}
\subsection{Mock data}
\begin{figure*}
    \centering
    \includegraphics[width=0.8\textwidth]{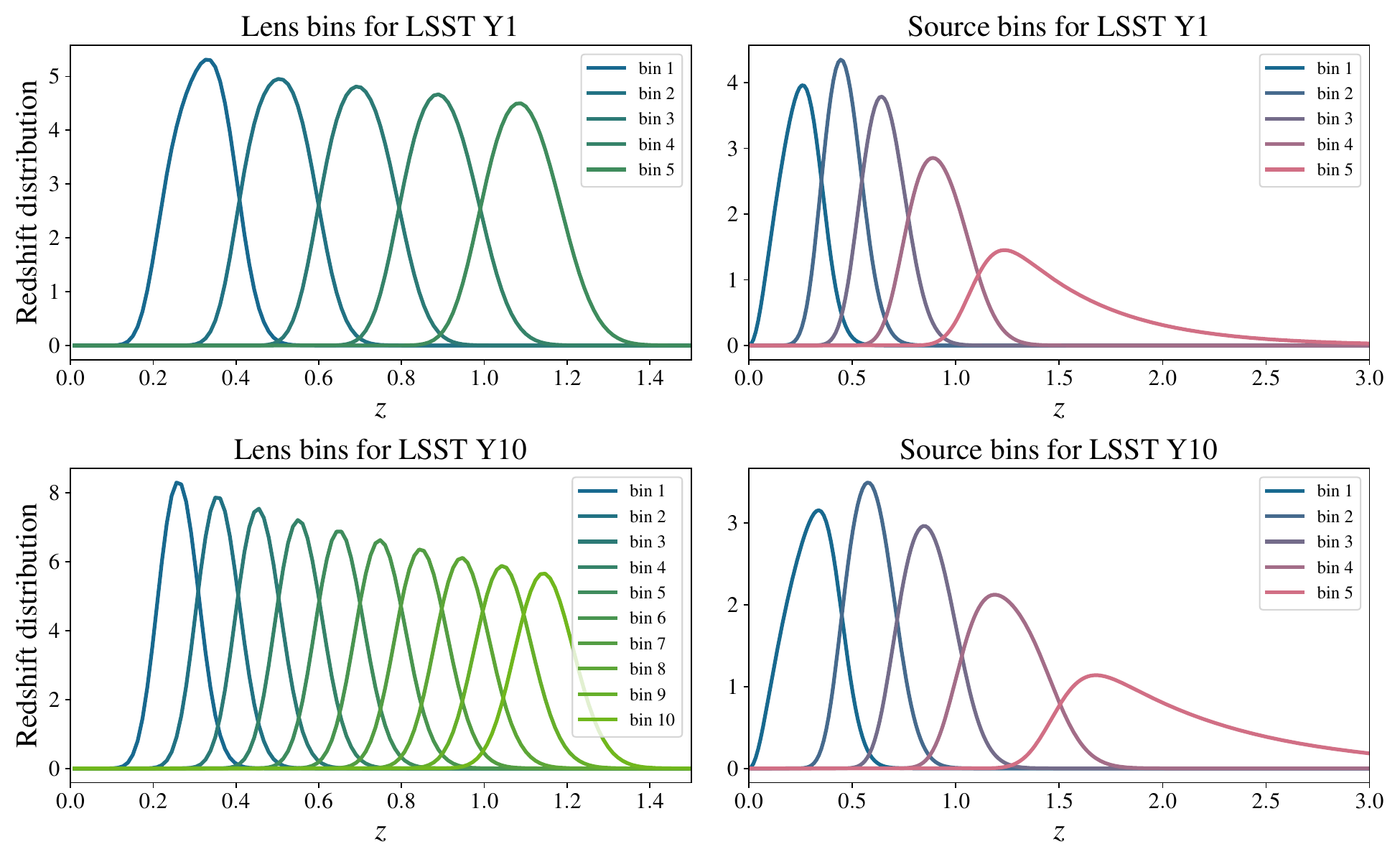}
    \caption{Redshift distributions of the tomographic bins for the lens (left) and source (right) samples as specified by the \citetalias{Mandelbaum2018} for LSST Y1 (top) and Y10 (bottom).}
    \label{fig:redshift_distribution}
\end{figure*}

For this work, we generated LSST Y1- and Y10-like mock cosmic shear and galaxy clustering spectra. All data vectors are noiseless and generated with theory pipeline discussed below assuming the redshift distribution based on the LSST-DESC science requirement document \citep{Mandelbaum2018}, hereafter \citetalias{Mandelbaum2018} in the range $z = [0.01, 3.0]$, as shown in  \autoref{fig:redshift_distribution}.
\footnote{\href{https://github.com/LSSTDESC/forecasting}{https://github.com/LSSTDESC/forecasting}}.
Our data vector consists of the cosmic shear, galaxy clustering auto, and galaxy--matter cross-spectra, which are created for 20 log-spaced bins in the range $\ell=[20,5000]$, assuming a sky fraction $f_{\rm sky}=0.43$ for both Y1 and Y10. 

We used two samples to perform the $3\times2$pt measurements: a source sample for shear and a lens sample for galaxy clustering. Following the \citetalias{Mandelbaum2018}, we assumed the source sample to be divided into 5 redshift bins in both Y1 and Y10, with an effective number density of $n^{s}_{\rm eff,Y1} = 10 \rm\,arcmin^{-2}$ and $n^{s}_{\rm eff,Y10} = 27 \rm\,arcmin^{-2}$. We use a lens sample divided into 5 redshift bins in Y1 and 10 bins in Y10, with $n^{l}_{\rm eff,Y1} = 18 \rm\,arcmin^{-2}$ and $n^{l}_{\rm eff,Y10} = 48 \rm\,arcmin^{-2}$. 
Also from the \citetalias{Mandelbaum2018}, the variance on the ellipticity (the shape noise per component) is taken to be $\sigma_e = 0.26$.

\subsubsection{Galaxy bias}
As previously mentioned, the HEFT bias model requires four bias parameters in each redshift bin. To get the bias fiducial values used for the mock data, we generated a redshift evolution of each parameter based on the fits in \citet{Nicola2024}, which were derived from the HSC DR1 survey \citep{Nicola2020}.

\subsection{Theory overview}
As discussed above, this work examines the interplay between baryonic feedback and bias parameters in a $3\times2$pt analysis designed to resemble the specifications of LSST Y1 and Y10. In general, the angular power spectrum $C_\ell^{XY}$ between two fields $X$ and $Y$ is defined as:
\begin{equation}
    \langle X_{\ell m} Y^*_{\ell'm'} \rangle \equiv C^{XY}(\ell) \delta_{\ell \ell'} \delta_{m m'}\,,
\end{equation}
where, in our case, the two tracers will be galaxy clustering (G) and shear ($\gamma$).

Under the Limber approximation \citep{Limber1953, LoVerde2008} and assuming a flat Universe cosmology, the 2D angular power spectra can be obtained as a projection of the corresponding 3D power spectra:
\begin{equation}
     C^{\rm XY}_{ij}(\ell) = \int d\chi \frac{W_{\text{X}_i}(\chi)W_{\text{Y}_j}(\chi)}{\chi^2} P_{\rm XY}\left(k=\frac{\ell+1/2}{\chi},z(\chi)\right) \, ,
    \label{eq:multitracer-cell}
\end{equation}
where the subscripts $i$ and $j$ indicate a specific redshift bin, $k$ is the wavenumber, $\ell$ is the 2D multipole moment and $\chi$ is the comoving distance, $W_{\text{X}_i}$ denotes the window function for probe $X$ and bin $i$, and $P_{\rm XY}(k,z)$ is the three-dimensional power spectrum for probe $X$ and $Y$. The window function for lensing tracers is given by:
\begin{equation}
    W_{\gamma_i}(\chi) = \frac{3H_0^2\Omega_{\rm m}}{2} \frac{\chi}{a(\chi)} \int_\chi^{\chi_h} d\chi^\prime  \, \frac{n_i(z)}{\bar{n}_i} \frac{dz}{d\chi^\prime} \frac{\chi^\prime - \chi}{\chi^\prime}\, ,
\end{equation}
where $H_0$ is the Hubble expansion rate today, $\Omega_{\rm m}$ is the fractional energy density of non relativistic matter, $a$ is the scale factor, $n(z)$ is the galaxy number density distribution with mean $\bar{n}$ and $\chi_h$ is the comoving distance to the horizon. The window function for galaxy clustering is instead given by:
\begin{equation}
    W_{G_i}(\chi) = \frac{n_i(z)}{\bar{n}_i}\frac{dz}{d\chi},
\end{equation}
where $n_i$  is the lens redshift distribution and $\bar{n}_i$ is the mean number density of lenses in the $i$-th tomographic bin.

The uncertainties of the data vector are modelled analytically, and we include only the Gaussian component of the $3\times2$pt covariance matrix (calculated with \mgl), since the other two -- the super-sample and non-Gaussian components -- are sub-dominant \citep{Joachimi2021}.

\subsubsection{Intrinsic alignments}
Cosmic shear measurements receive two additional contributions, known as intrinsic alignment (IA), to the one from the pure gravitational lensing effect (cosmic shear): one contribution (II) from the coherent alignment induced by the underlying tidal field on physically close galaxies and one coming from the cross-correlation with cosmic shear signal ($\gamma$I) \citep{Catelan2001, Hirata2004, Paopiamsap2024}. The observed shear angular power spectrum is therefore:
\begin{equation}
    C^{\rm obs}_{ij}(\ell) = C^{\rm \gamma\gamma}_{ij}(\ell) + C^{\rm \gamma I}_{ij}(\ell) + C^{\rm II}_{ij}(\ell) \,.
\end{equation}

In this work the intrinsic alignment effect is modelled using the redshift-dependent nonlinear alignment model (NLA-z) \citep{Bridle2007, Joachimi2021}, 
which assumes in Fourier space the matter-intrinsic and the intrinsic-intrinsic to be written as \citep{Lamman2024}:
\begin{align}
    \begin{split}
        P_{\rm mI}^{\rm NLA}(k,z) &= f_{\rm IA}(z) P_{\rm mm}^{\rm NL}(k,z) \, , \\
        P_{\rm II}^{\rm NLA}(k,z) &= f^2_{\rm IA}(z) P_{\rm mm}^{\rm NL}(k,z) \, ,  
    \end{split}
\end{align}
where $P_{\rm mm}^{\rm NL}$ is the nonlinear matter power spectrum, while $f_{\rm IA}$ is defined as
\begin{equation}
    f_{\rm IA}(z) = -A_1 C_1 \rho_c \, \frac{\Omega_{\rm m}}{D(z)} \left( \frac{1+z}{1+z_{\rm piv}} \right)^{\eta_1} \, , 
    \label{eq:IA}
\end{equation}
where $C_1$ is a normalization constant, $\rho_c$ is the critical density today, $D(z)$ is the growth factor, and $z_{\rm piv}$ is a pivot redshift \citep{Joachimi2021} which we fix to $0.62$. This model has been adopted in many analyses, see for example \citet{Krause2016}. The model has only two free parameters: the dimensionless amplitude $A_1$, which sets the overall strength of the contamination, and $\eta_1$ which controls its redshift dependence and captures the sensitivity of IA to the galaxy sample.
Our chosen fiducial IA parameter values are $A_1=0.36$ and $\eta_1=1.66$, taken from \citet{Secco2022, MacMahon-Geller2024}.

\subsubsection{Scale cuts}
We choose to explore the impact of scale cuts on the galaxy-galaxy correlation function, while keeping the lensing cuts fixed to $\ell_{\rm max}= 2000$ in each bin. We chose this value following the \citetalias{Mandelbaum2018} guidelines, which also represent a safe choice given the limitations of \bacco. For simplicity, the same galaxy-galaxy scale cuts are also applied to the galaxy-shear cross-correlation function
\footnote{Since galaxy--galaxy and galaxy--lensing probe different scales, in a complete analysis they should have customised scale cuts.}
. In particular, to specify the retained region in $k$-space, we define a maximum scale, $\kmax$, which is then mapped to the corresponding $\ell_{\rm max}$ as
\begin{equation}
    \ell_{\rm max} = \kmax d \,,
\end{equation}
where $d$ is the angular diameter distance $d = \chi_{\rm max}/(1+z_{\rm mod})$, with $z_{\rm mod}$ the redshift corresponding to the peak of the kernel for the source redshift distribution and $\chi_{\rm max} = \chi(z_{\rm mod})$ is the corresponding comoving distance.

\subsection{\mgl ~pipeline} \label{Subsec:mgl}
\mgl{} (Modified Gravity Lensing)\footnote{\href{https://mglensing.readthedocs.io/en/latest/}{https://mglensing.readthedocs.io}} is a Bayesian inference code for different photometric probes that performs MCMC analysis for Stage IV surveys. Without inclusion of systematic effects on large scales, it focuses on systematics on small and nonlinear scales for standard and extended cosmologies. The cross- and auto-correlated power spectra are calculated using emulators, accelerating computation of the theoretical prediction. Both the mock data set and analytical Gaussian covariance are computed at some fiducial cosmology, in the subsequent analysis the posterior distribution is explored with \textsc{nautilus} sampler \citep{Lange2023}, with $n_{\rm eff}=5000$, $n_{\rm live}=1000$ \footnote{We tested that higher values of $n_{\rm live}$ (3000 for Y1 and 7000 for Y10) have no impact on the posterior distributions}. 
We discuss the Gaussian-only covariance assumption in Appendix \ref{Appendix:tot_cov}.
For \bacco ~emulator extrapolation routines for $z>1.5$ and $k>0.7\,h/$Mpc (for HEFT expansion) and $k>5\,h/$Mpc (for nonlinear matter power spectrum and baryonic feedback) are implemented.  In this paper, for the extrapolation in redshift, we assume the linear power spectrum computed with \bacco. A comparison with the extrapolation via the nonlinear power spectrum computed with HMcode2020-emulator\footnote{\href{https://github.com/MariaTsedrik/HMcode2020Emu/tree/main}{https://github.com/MariaTsedrik/HMcode2020Emu}} showed no visible impact on our constraints. For the extrapolation in scale, we apply the power law extrapolation. Extrapolating with the constant value at the highest valid $k$-value also led to the same constraints. 

A comparison with the DESC Core Cosmology Library pipeline \citep{Chisari2019} for the model with \bacco ~emulator used throughout this work is discussed in Appendix \ref{Appendix:ccl}.\footnote{Extended comparisons are provided in \href{https://github.com/MariaTsedrik/MGLensing/blob/main/tutorials/mgl_vs_ccl_hmcode.ipynb}{notebook 1} and \href{https://github.com/MariaTsedrik/MGLensing/blob/main/tutorials/mgl_vs_ccl_bacco.ipynb}{notebook 2}.}

\subsection{Figures of merit and bias}
We quantify the results of fitting (in some cases mismatched) models to mock data with two quantities.
The figure of bias (FoB) quantifies the relative separation of the measured parameters from their fiducial values in terms of the variance of the posterior distribution. It is defined as \citep{Shapiro2009, Joudaki2009}:
\begin{equation}
    \mathrm{FoB} = \sqrt{(\theta_\mathrm{fid} - \bar{\theta}) S^{-1} (\theta_\mathrm{fid} - \bar{\theta})} \, ,
\end{equation}
where $\bar{\theta}$ and $\theta_\mathrm{fid}$ are vectors labelling, respectively, the posterior averages and fiducial values of the parameters we want to measure, and $S = \mathrm{cov} (\theta)$ is the covariance matrix of the parameters calculated from the chains.
 
We calculate the 68\% - 95\% thresholds for the FoB by assuming the posterior distribution of the parameters of interest to be Gaussian. In other words, the confidence intervals for FoB are calculated by direct integration of a two-dimensional Gaussian over an ellipse between $\pm$ FoB and equating the integral to the 68th and 95th percentile thresholds, which in a two parameter case results in FoB equal to 1.52 and 2.49, respectively. Note that even for correct analysis -- \textit{i.e.}, where the same model is assumed both in the mock data and in the likelihood analysis -- the FoB can be non-zero due to projection effects \citep{Hadzhiyska2023}.

We quantify the constraining power of a considered model with respect to the parameters varied in the fit with the figure of merit (FoM). The FoM is defined as the inverse of the volume of the 68\% contours of the parameters, effectively giving a global measure of the inverse size of the uncertainties obtained on the parameters \citep{Wang:2008zh}:
\begin{equation}
    \mathrm{FoM} = \frac{1}{\sqrt{\mathrm{det}(S)}} \, .
\end{equation}	
This quantity also assumes the parameter distribution to be Gaussian.

\begin{figure*}
    \centering
    \includegraphics[width=0.45\textwidth]{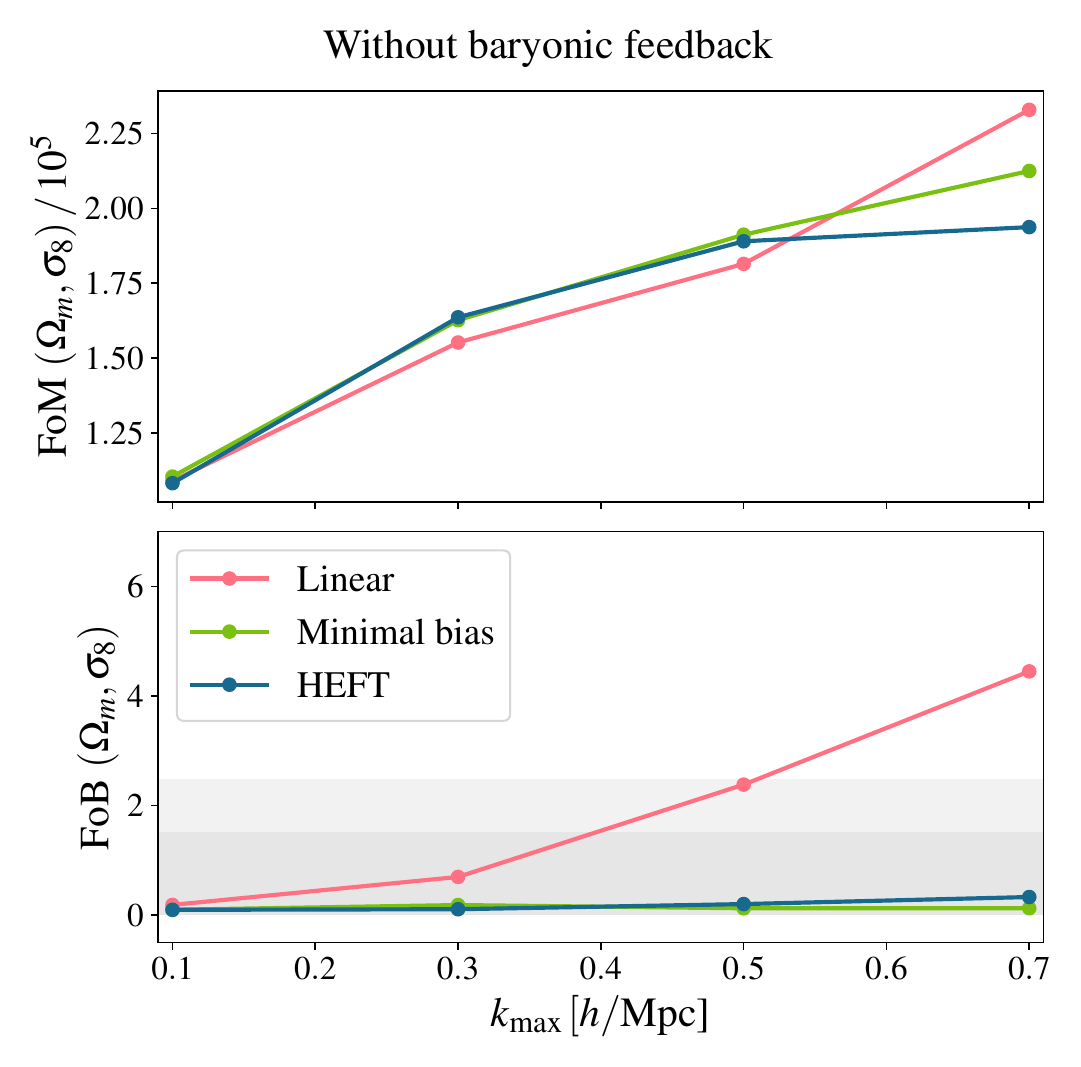} \quad 
    \includegraphics[width=0.45\textwidth]{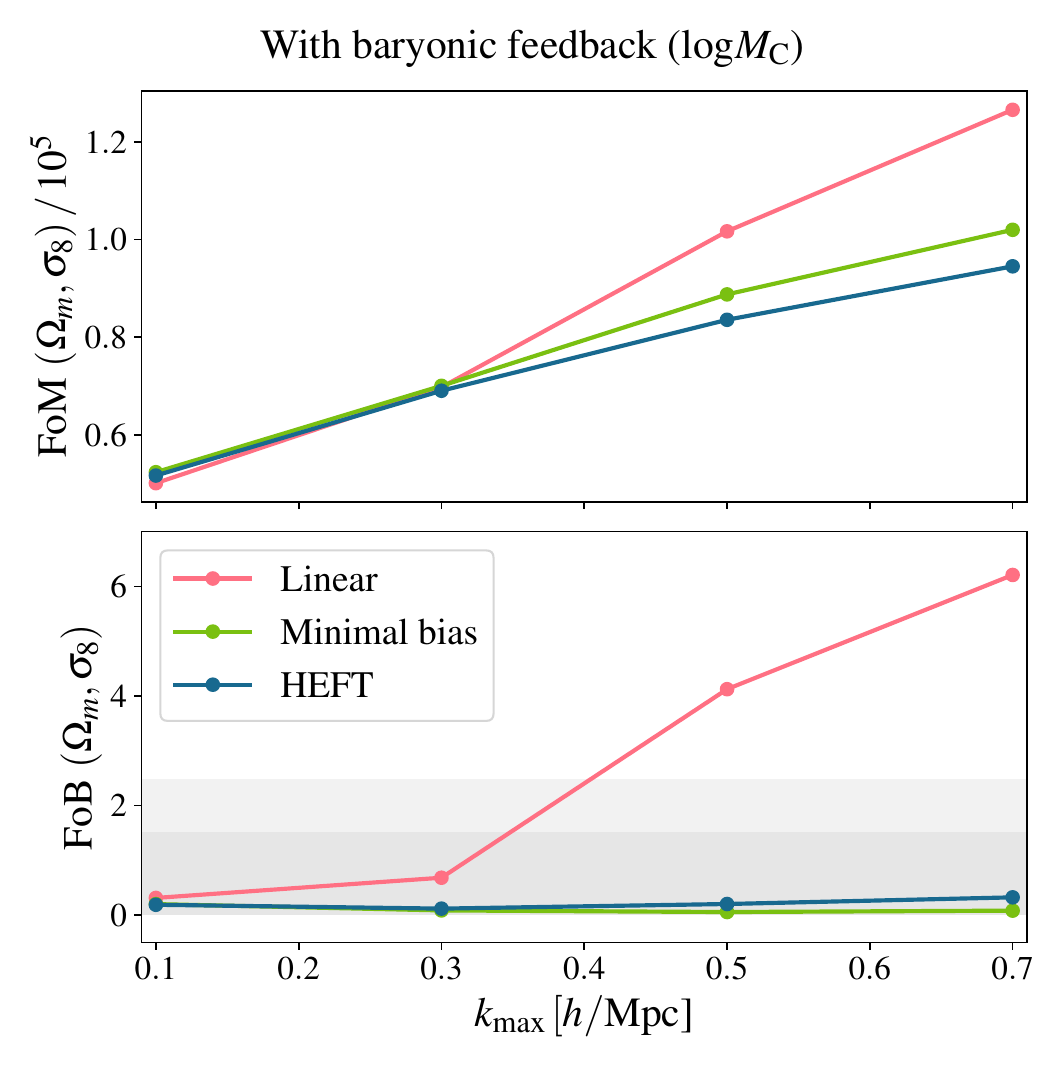}
    \caption{Figure of merit (top) and figure of bias (bottom) for $\Omega_{\rm m}$ and $\sigma_8$ at different scale cuts for LSST Y1. Left panel: without baryonic feedback; Right panel: baryonic feedback is explicitly included in shear and modelled as in \autoref{eq:Pgg_Pgm_zennaro} in galaxy clustering, with the only free parameter being $\logMc$. All cases consider HEFT as fiducial model for the data, while we used the linear bias (pink line), minimal bias (green line) and HEFT (blue line) models in the analysis. The gray shaded areas in the lower panels represent the 1 and 2-$\sigma$ thresholds with respect to the fiducials.}
    \label{fig:fom_fob_bias}
\end{figure*}

\section{Results}\label{Sec:Results}
In this section, we present the results of this work, with particular focus on the interplay between baryonic and bias parameters. To this end and to test the robustness of model misspecification, we analyse scenarios where the models assumed in the mock data and in the subsequent likelihood analysis do not always match. The specific choices are stated for each case.
In this work we chose to keep the photo-$z$ error fixed. Note that, unless specified, all figures in this section refer to LSST Y1.

\subsection{Bias model simplification}
\label{sec:scalecuts}
As previously mentioned, describing the nonlinear behaviour of large-scale structure (LSS) requires complex modelling with the introduction of various nuisance parameters, some of which are degenerate with each other, as we will discuss in more detail in this section. 

In this work we take advantage of our lack of knowledge on how to constrain these nuisance parameters (\textit{e.g.}, which would require additional astrophysical and cosmological probes beyond the $3\times2$pt analysis considered here) by fixing some of them. This will reduce the complexity of the model avoiding any loss of accuracy or the need to discard more data through scale cuts.

\subsubsection{Simplifying bias (without baryons)}
We begin by investigating whether the number of bias parameters can be reduced in the absence of baryonic effects. 

A mock dataset was created assuming a nonlinear power spectrum with a suppression factor $S=1$, and with HEFT bias parameters set to their fiducial values provided by the DESC collaboration (see \autoref{tab:fiducials}). 
The simplest test we can perform is then to adopt a linear bias model to analyse these data. The pink line in \autoref{fig:fom_fob_bias} shows a comparison of the obtained $\Omega_{\rm m}$ and $\sigma_8$ FoM and FoB across different scale cuts to a reference case (blue line) that employs the full HEFT model. As expected, assuming a linear bias model introduces a bias in the results: in FoB, the line exits the 95th percentile region for scale cuts at around $k=0.5\,h/$Mpc. 
Depending on specific requirements, a linear bias model with cuts $k\lesssim0.3\,h/$Mpc might prove acceptable; this corresponds to a $\sim0.5\sigma$ bias in $\Omega_{\rm m}$ and $\sigma_8$. This could, however, be risky. It depends on our specific configuration of broad flat priors on $\Omega_b$, $h$, and $n_{\rm s}$, and the latter two are seen to be biased in \autoref{fig:corner-b1-heft-fix-kmax0.3}. Adding additional data sets or tighter priors would almost certainly lead to more clearly unacceptable biases in $\Omega_{\rm m}$ and $\sigma_8$. Hence, in these scenarios we recommend applying a conservative scale cut of $\kmax=0.1\,h/$Mpc.

A straightforward way to simplify the model described in \autoref{eq:galaxy_bias_expansion} is to fix $\bs$ and $\blap$ to zero in the analysis, we will refer to this setup as ``minimal bias model'' \citep{Nicola2024}. 
This simplification is also in agreement with the so-called local-in-matter-density (LIMD) Lagrangian bias, which assumes the connection between the tracer overdensities at early times to be solely a function of the underlying local matter density \citep{Desjacques2018}.
The corresponding result is shown by the green line in \autoref{fig:fom_fob_bias}, where we find that there is no significant difference with the reference case. This can be explained by the physically motivated choice of the fiducial values for $\bs$ and $\blap$ being close to zero (see \autoref{tab:fiducials})
\footnote{In our tests we found that adopting extremely high values of the higher-order bias parameters as fiducial values (for instance $\blap \sim 1.8 - 2.9$) in the synthetic data leads to highly non-Gaussian distributions in $h$, $A_{\rm s}$ and $n_{\rm s}$. This scenario also leads to higher values of the $\Omega_{\rm m}$ and $\sigma_8$ FoM and FoB. As found in \citet{Nicola2024} for Y10, we warn that this result is sample and analysis dependent and one should be cautious in choosing to use the minimal bias model.}
In the full HEFT case, with increasing $k_{\rm max}>0.5\, h/$Mpc the constraining power is gained only in higher-order bias parameters, hence the plateau in FoM. The slightly higher FoB of the full HEFT case is the result of non-Gaussian posterior distributions in $b_2$, $\bs$ and $\blap$ in some of the redshift bins (see \autoref{fig:corner-heft-fix-kmax0.7} in Appendix \ref{appendix-extra-figs-tables}), as well as unconstrained $\Omega_{\rm b}$, $h$ and $n_{\rm s}$. We also checked that by fixing these three cosmological parameters to their fiducial values, the linear bias case has a higher FoM than the others -- unlike the case shown in the plot -- as we would expect since it is the case with least number of free parameters.

\subsubsection{Simplifying bias (with baryons)}\label{Subsubs:simplifyingbias} 
A similar analysis could, in principle, be performed for the full set of the baryonic parameters. However, it is well known in the literature that the 7 parameters required by the baryonification approach are mostly unconstrained, except for $\logMc$. Studies have also demonstrated that one or two parameters are sufficient to capture the main effects of the baryonic feedback at a given redshift \citep{Arico2021, Schneider2025, Wayland2025}. Moreover, as in this work we solely focus on different galaxy clustering scale cuts -- which are always more stringent than shear cuts -- it is reasonable to vary only the $\logMc$ parameter in our analysis
\footnote{All analyses in this work which include baryonic feedback vary only $\log M_c$, while keeping all other baryonic parameters fixed.}.
We discuss the validity of this assumption in Appendix \ref{Appendix:baryons}.
The baryonic feedback is explicitly added in shear power spectrum by multiplying the suppression factor by the nonlinear matter-matter power spectrum, and modelled as in \textit{Zennaro24} in galaxy clustering and galaxy--galaxy lensing. 

As before, we evaluate the FoM and FoB for $\Omega_{\rm m}$ and $\sigma_8$ (see right panel in \autoref{fig:fom_fob_bias}) to quantify the loss in constraining power when the baryonic parameter is added to the analysis. We find that the FoM values are decreased by approximately half compared to the previous cases; nevertheless, the overall trend remains unchanged. 
With the minimal bias model extending the scale cut from $\kmax=0.1\,h/$Mpc to $0.7\,h/$Mpc yields an FoM gain of approximately a factor of two, compensating for the comparable loss in constraining power induced by the baryonic parameter. This demonstrates once more that our cosmological constraints are not weakened by adopting a bias model more complex than linear -- in fact, they benefit from it.

In summary, in both cases -- with and without baryonic feedback -- a linear bias model is a good approximation which gives unbiased constraints for $\Omega_{\rm m}$ and $\sigma_8$ when using $k\lesssim0.1\,h/$Mpc (FoB $< 0.5\sigma$).

Additionally, we find that adopting the hybrid-perturbative approach for galaxy bias does not degrade cosmological constraints compared to using only linear bias and linear scales. We also find that the precision of cosmological constraints continuously improves with inclusion of smaller scales, up to the validity limit of HEFT $\kmax=0.7\,h/$Mpc. This relative gain in the constraining power remains even with the inclusion of the baryonic parameter in the analysis, although the absolute constraining power is degraded.
We also show that this complex model might be further simplified via omission of $\bs$ and $\blap$ parameter, as the minimal bias model does not lead to biased cosmological estimates even when extending the analysis to $\kmax=0.7\,h/$Mpc. We therefore advocate for the inclusion of quasi-nonlinear scales in Stage IV analyses, noting that a minimal bias parameterization may be sufficient in practice. We emphasize, however, that this conclusion holds for the specific fiducial bias parameter values and cosmology adopted in our analysis.

\begin{figure*}
    \centering
    \includegraphics[width=\textwidth]{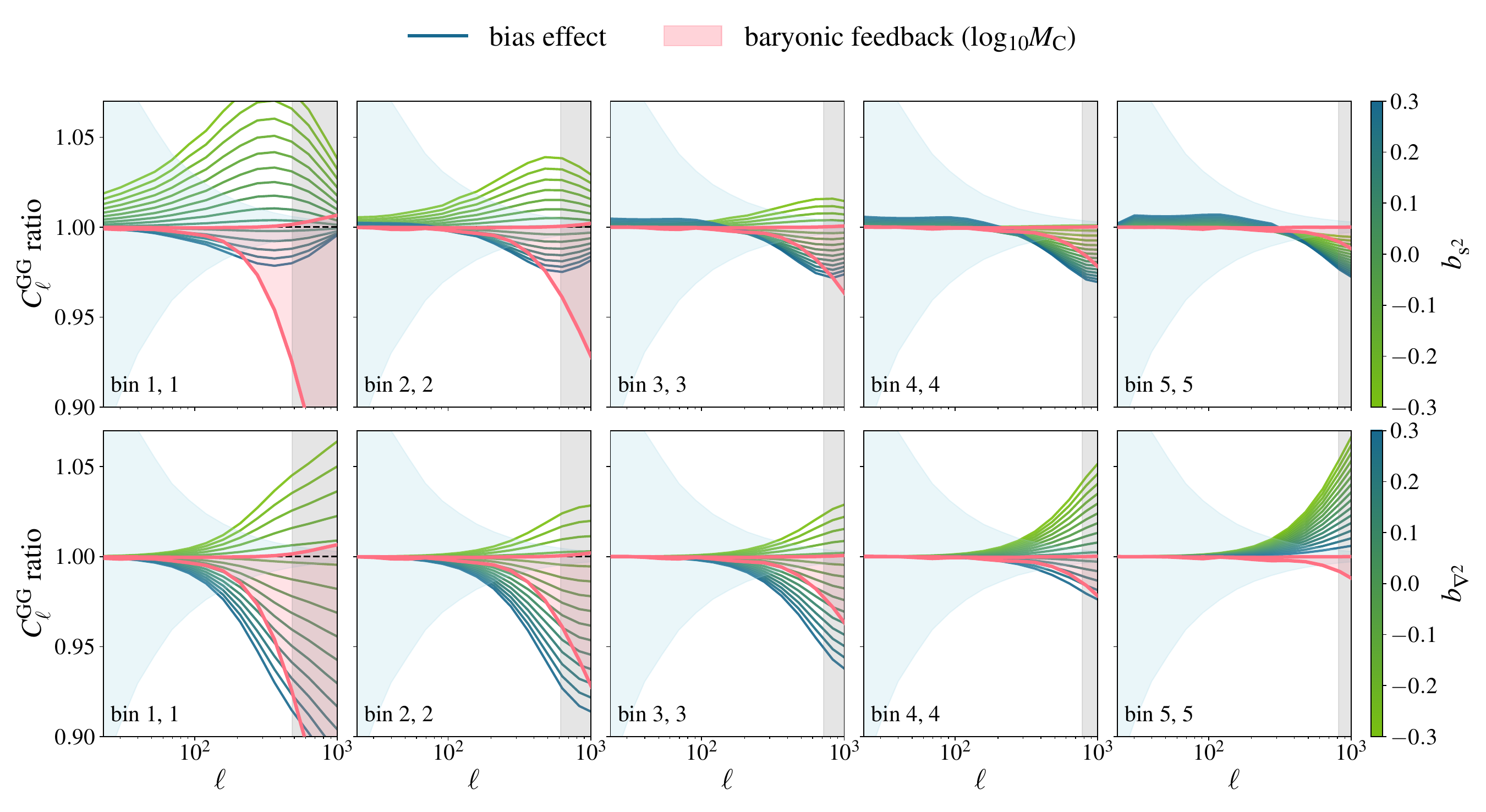}
    \caption{Galaxy-galaxy auto-correlation angular power spectra showing how bias parameters mimic the effect of baryonic suppression. In pink we show the suppression factor $S$, in particular, the two lines and the shaded region in between correspond to the power spectra range with baryonic suppression with $\logMc$ between 9 and 15.
    The gradient-colored lines include higher-order bias contributions, modelled with HEFT, and no baryonic suppression factor. 
    All ratios with varying values of the higher-order bias parameters are normalised to the fiducial galaxy-galaxy angular power spectrum used in the mock data (see \autoref{tab:fiducials} for the numerical values of $\bs$ and $\blap$ in each redshift bin).
    The gray areas indicate the scale cut $\kmax = 0.7\, h/$Mpc, beyond which the HEFT model is no longer valid.}
    \label{fig:rainbow_plot}
\end{figure*}

\begin{figure*}
    \centering
    \includegraphics[width=\textwidth]{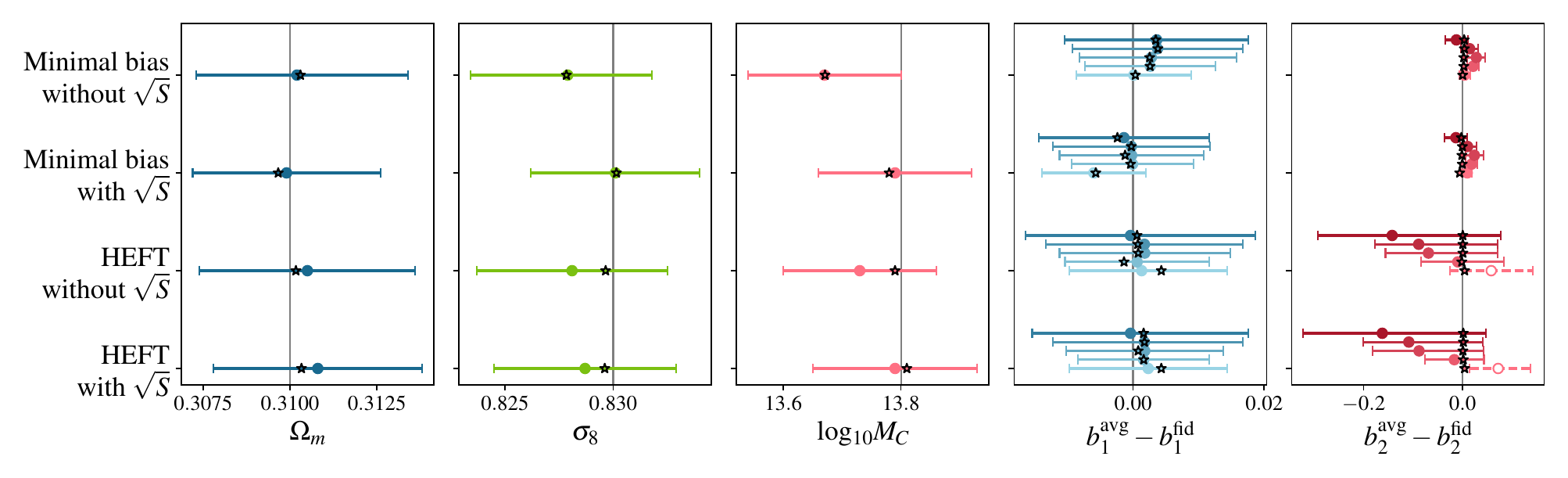}
    \caption{Mean value and $1\sigma$ error of $\Omega_{\rm m}$, $\sigma_8$, $\logMc$ and the first two bias parameters $b_1$ and $b_2$ for four different modelling assumptions: full HEFT bias with and without baryonic suppression included in GGL, and same two scenarios but using the minimal bias approach instead of full HEFT. All cases consider the same mock data with HEFT as bias model and with baryonic feedback. The grey vertical lines correspond to the fiducial values of the parameters and the errorbars to the standard deviation. The bias parameters are shown for the five redshift bins (from light to dark as going further in redshift). We marked $b_2^1$ with empty marker and dashed errorbars to highlight that its posterior distribution is bimodal. The black star symbols show the best fit values, obtained with \texttt{iminuit}.}
    \label{fig:avg_std_y1_sqrtS}
\end{figure*}

\subsection{Baryon-Bias Interplay}

In this section we test whether the galaxy bias expansion — especially its higher-order contributions, $\bs$ and $\blap$ — can effectively mimic the suppression in the power spectrum associated with baryonic feedback processes (and vice-versa). While we focus on LSST Y1, we have verified that analogous conclusions hold for LSST Y10.

The depiction of our intuition that baryonic suppression and higher-order bias parameters might exhibit similar behaviour in galaxy clustering $C_\ell$ is shown in \autoref{fig:rainbow_plot}. There we compare two cases. 
The shaded pink regions in the figure denote the prior ranges of baryonic suppression considered in our analysis. This range is obtained by varying $\logMc$ within its prior bounds, and computing the ratio of galaxy-galaxy angular power spectra derived from $P_{\rm gg}=b_1^2P_{\rm NL}S$ and $P_{\rm gg}=b_1^2P_{\rm NL}$, respectively for the numerator and denominator. We use this naive modelling approach to isolate the effect of baryonic suppression and illustrate potential changes to the shape of $C_\ell^{\rm GG}$.
In the second scenario, shown by the gradient-colored curves, we assume no baryonic suppression (i.e., $S_{\rm mm}=1$), and instead allow for non-zero values of $\bs$ and $\blap$.
All ratios with varying values of the higher-order bias parameters are normalised to the fiducial galaxy-galaxy angular power spectrum used in the mock data (\textit{i.e.}, with neither baryonic feedback nor higher-order bias).
The plot clearly shows that for a combination of suitably chosen values of $\bs$ and $\blap$, the resulting power spectrum closely matches the reference one within uncertainties (represented by the blue shaded areas in the plot) for multipoles corresponding to $k_{\rm max} \lesssim 0.7 \,h$/Mpc (gray shaded areas) beyond which the HEFT approach is no longer valid. We quantify this comparison in more detail in Appendix~\ref{Appendix:quantify_rainbow}.

\subsubsection{Galaxy-galaxy lensing} 
We next tested whether an incorrect model of baryonic feedback in the galaxy-galaxy lensing component leads to biased cosmological parameters. \citet{Zennaro2024} showed that equation \autoref{eq:Pgg_Pgm_zennaro} accurately describes the 3D $P_{\rm gm}$ in the FLAMINGO simulations well up to $k_\mathrm{max}=0.7 h/$Mpc at $z=0$ and $z=1$, but also showed that neglecting the $\sqrt{S}$ term in that equation does not lead to cosmological parameter biases. 

Here we extend the tests in \citet{Zennaro2024} by working with observable $C_\ell$'s instead of matter and galaxy power spectra directly. The consequent changes to scale and redshift sensitivity, and complex posterior degeneracies, could complicate their findings. This check also acts as a proxy for wider uncertainty in $P_{\rm gm}$ modelling. We also wish to test if their result is sensitive to the use of the simpler minimal bias model used in \citet{Nicola2024}, which neglects two HEFT parameters per bin.

To answer these questions, we construct a LSST Y1 mock data, assuming the HEFT model for bias and baryonic effects included as \autoref{eq:Pgg_Pgm_zennaro} in galaxy-galaxy (GG) and galaxy-galaxy lensing (GGL). We then analyse these mocks under four different modelling assumptions: full HEFT bias with and without baryonic suppression included in GGL, and the same two scenarios but using the minimal bias approach instead of the full HEFT.
In all four analyses, we fixed IA parameters and all cosmological parameters except for $\Omega_{\rm m}$ and $\sigma_8$ to mimic the analysis in \citet{Zennaro2024}.

The recovered mean values of $\Omega_{\rm m}$, $\sigma_8$, $\logMc$ and the first two bias parameters in each bin, $b_1^i$ and $b_2^i$, for the four analysis setups are shown in \autoref{fig:avg_std_y1_sqrtS}; the bottom row in that figure has the same modelling prescription as the mock data, so shifts there are entirely projection biases.
We find that cosmological parameters are consistent with their fiducial values (vertical grey lines in the plot) in all scenarios. 

As expected, the most significant difference appears in the baryonic parameter $\logMc$ parameter, which, when the suppression factor is set to $S=1$ in GGL, is effectively constrained by the shear signal (with $S\neq1$) only. This leads to a change in its inferred value but, as in \citet{Zennaro2024}, without biasing the cosmological parameters.

All linear bias values are consistent with the respective fiducials in each redshift bin (within $1\sigma$ error).
The parameter that shows the biggest change in value and errorbar is $b_2$. For this parameter, as expected, the errors are significantly smaller in the case of minimal bias compared to the HEFT scenario because of the smaller number of free parameters. The peak of the posterior is also shifted in most of the redshift bins as this parameter is trying to include the effect of the higher order bias parameters which are here set to zero. Note that we indicated $b_2^1$ with empty markers and dashed errorbars to highlight that its posterior distribution is bimodal, similarly to what \autoref{fig:corner-heft-fix-kmax0.7} in Appendix \ref{appendix-extra-figs-tables} shows for a different setup.

\citet{Zennaro2024} found, in a similar analysis, a bias in the recovered galaxy bias parameters, especially $b_2$, caused by the incorrect baryon model. We see no such bias here, because the projection from $P(k)$ to $C_\ell$ results in wider posteriors and washes out the difference. A direct comparison with the minimal bias cases, which have error bars of a similar order of magnitude, is also not meaningful, since in our minimal bias model $b_2$ partially absorbs the effect of the higher-order bias parameters, which are instead explicitly included in the \citet{Zennaro2024}.

Overall, these results indicate that the bias parameters are flexible enough to mimic the effect of baryons in galaxy–galaxy lensing without including a significant bias in the cosmological parameters, even when baryonic corrections in GGL are omitted in the analysis, as found in \citet{Zennaro2024}. However, the minimal bias model might lead to shifts in baryonic parameters when baryons are not accounted for in GGL. This conclusion also holds when all systematic and cosmological parameters are varied in the analysis. The only difference we observed is in the minimal bias case without $\sqrt{S}$ in GGL: with respect to the fiducial value, $\sigma_8$ is shifted towards lower values by $\sim$$1\sigma$. Based on the full contour plot (not shown) this is potentially due to the weak constraints and degeneracies with $\Omega_{\rm b}$, $n_{\rm s}$ and $h$.
\vspace{18pt}

\subsection{Bias model simplification in LSST Y10}
\begin{figure}
    \centering
    \includegraphics[width=\linewidth]{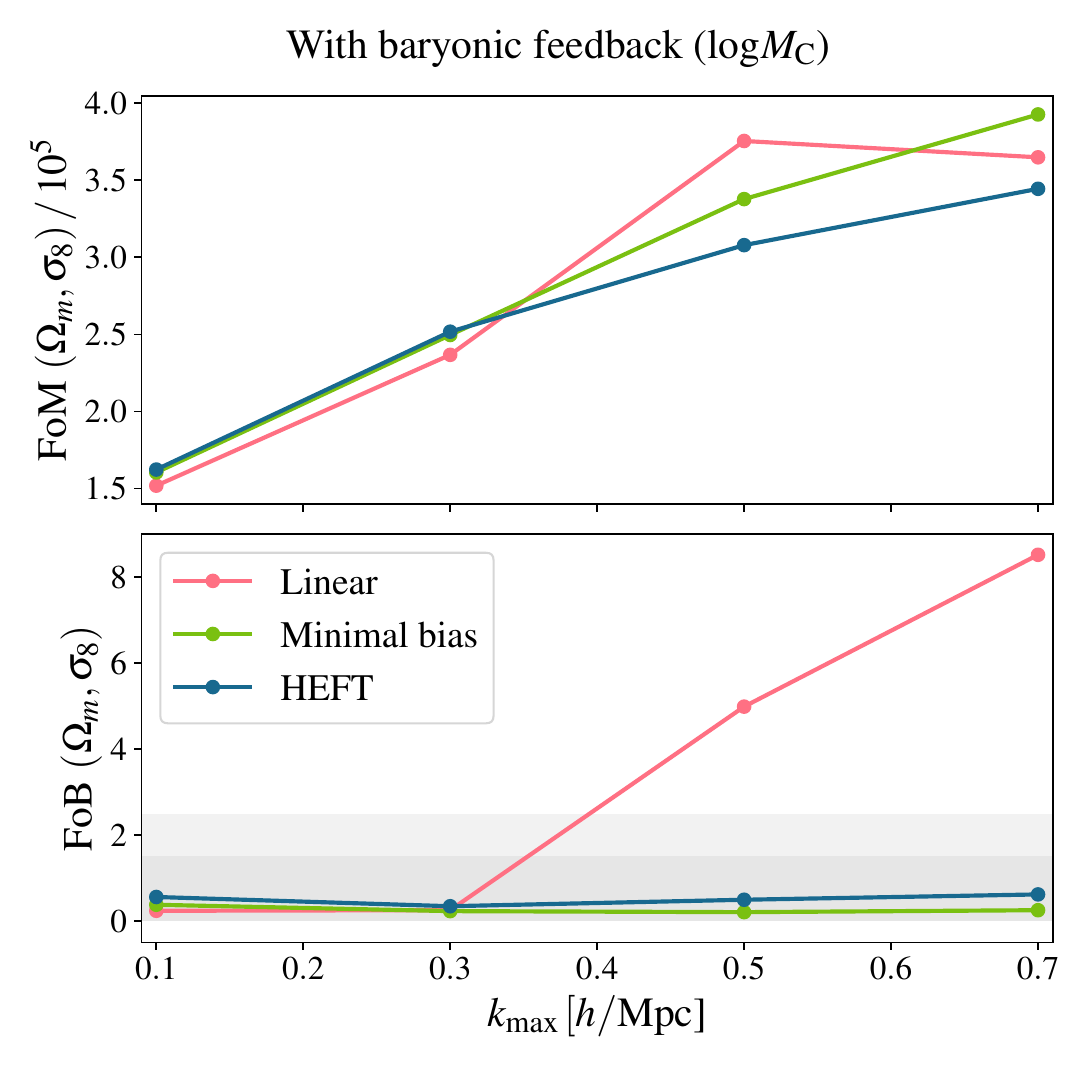}
    \caption{Figure of merit (top) and figure of bias (bottom) for $\Omega_{\rm m}$ and $\sigma_8$ at different scale cuts for LSST Y10. Baryonic feedback is explicitly included in shear and modelled as in \autoref{eq:Pgg_Pgm_zennaro} in galaxy clustering and cross-correlation, with the only free parameter being $\logMc$. All cases consider HEFT as fiducial model for the data, while we used the linear bias (pink line), minimal bias (green line) and HEFT (blue line) models in the analysis. The gray shaded areas in the lower panels represent the 1 and 2$-\sigma$ thresholds with respect to the fiducials.}
    \label{fig:fom_fob_y10}
\end{figure}

LSST is planned to run for 10 years, gaining more data and information each year. To assess the impact of the full survey on constraining cosmological parameters, we carried out a set of a few forecast cases for Y10, analogous to what was done for Y1.

The number of lens bins is expected to increase from around five in Y1 to around ten in Y10; using the full bias model would therefore require 40 bias parameters. For this reason, we again investigate whether the number of bias parameters can be reduced in the analysis. Following the same procedure adopted for LSST Y1, we generated a mock dataset assuming a nonlinear power spectrum with baryonic suppression modelled as in \autoref{eq:Pgg_Pgm_zennaro} and the HEFT bias model. We then analyse the dataset using three different bias models: linear bias, minimal bias and HEFT. The resulting FoM and FoB for $\Omega_{\rm m}$ and $\sigma_8$ as a function of scale cuts are shown in \autoref{fig:fom_fob_y10}. 

As expected, there is a general improvement in the FoM of Y10 compared to Y1. Particularly, at $\kmax=0.7\,h/$Mpc for HEFT and the minimal bias model, the FoM of Y10 is approximately 1.8 times that of Y1, illustrating the precision and information that LSST can extract from small scales as it accumulates data over the course of the survey. As our analysis is simplified, we expect the actual difference between LSST Y1 and Y10 to be higher, especially when introducing systematic effects we ignored, such as the photo-$z$ error.

Similarly to the Y1 case, linear bias provides a good approximation (FoB $<0.5$) only for $k\leq0.3\,h/$Mpc, while the minimal bias model remains valid up to the smallest scales considered in this analysis. The slight increase in FoB for the HEFT scenario on small scales, which results in a larger FoB than for the minimal bias, arises from projection effects due to the greatly expanded parameter space.
Note that the HEFT scenario at $\kmax=0.7\,h/$Mpc has FoB $\simeq0.6\sigma$. Looking at the full posterior distribution, this can be explained (similar to the Y1 analysis) as the result of non-Gaussian posterior distributions in the bias parameters in some of the redshift bins.

Based on these results, we might expect the minimal bias model to be safe also for LSST Y10 analyses. However, as we will see in the next section, while $\Omega_{\rm m}$ and $\sigma_8$ are robust in such an analysis, extension parameters and alternative choices of parameters can be more fragile.

\subsection{Massive neutrinos}
\label{sec:neutrinos}
\begin{figure}
    \centering
    \includegraphics[width=\linewidth]{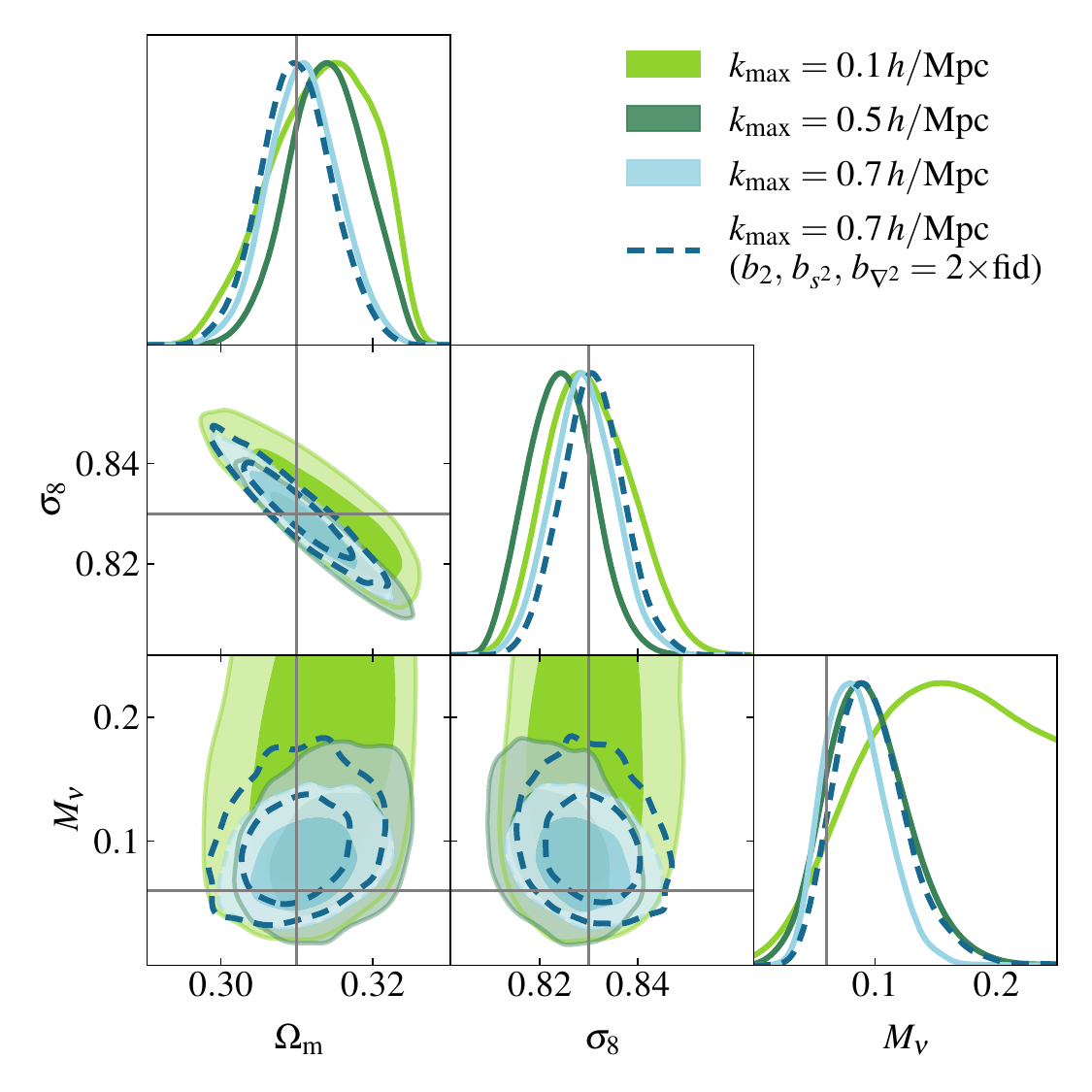}
    \caption{The 68\% and 95\% 2D confidence regions and posteriors for some of the parameters in the analysis at different scale cuts for LSST Y1. We assumed HEFT in the mock data and the minimal bias model in the analysis, both with baryonic feedback and massive neutrinos (free $\logMc$ and $M_\nu$). The solid contours consider the fiducial values as in \autoref{tab:fiducials}, while the dashed blue contour consider the fiducial values in the mock data of $b_2$, $\bs$ and $\blap$ to be twice as those.}
    \label{fig:Y1_massive_neutrinos}
\end{figure}

In this section, we want to analyse whether LSST will be able to clearly detect a non-zero total neutrino mass. To this end, we focus on its constraining power across different scales by varying $M_\nu$ as well as the cosmological parameters and $\logMc$, with fiducial values as in the standard cosmological model (see \autoref{tab:fiducials}). 

\begin{figure}
    \centering
    \includegraphics[width=\linewidth]{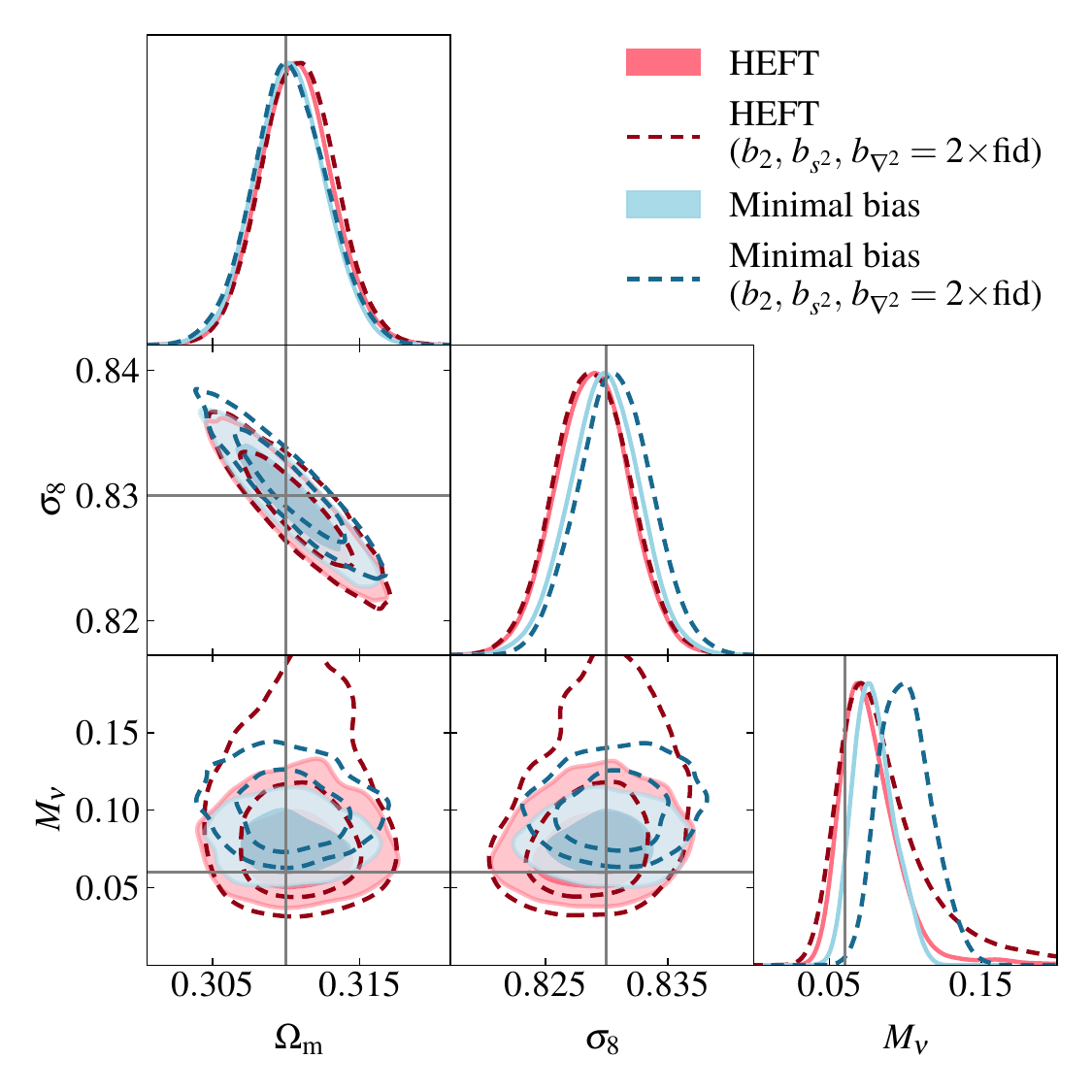}
    \caption{The 68\% and 95\% 2D confidence regions and posteriors for $\Omega_{\rm m}$, $\sigma_8$ and $M_\nu$ in the Y10 analysis at $\kmax=0.7\,h/$Mpc for four different scenarios using full HEFT (pink and red) and minimal bias (blue and dark blue) in the modelling. The solid contours consider the fiducial values as in \autoref{tab:fiducials}, while dashed contours represent the result of the same analysis, but where the fiducial values of $b_2$, $\bs$ and $\blap$ are set to a higher value (specifically, twice as in \autoref{tab:fiducials}).
    }
    \label{fig:Y10_massive_neutrinos}
\end{figure}

The solid contours in \autoref{fig:Y1_massive_neutrinos} show the Y1 results, obtained assuming the minimal bias model and different scale cuts. These indicate that a potential neutrino detection requires extending the analysis to relatively small scales $k\geq0.5\, h/$Mpc, implying that linear bias models are insufficient and perturbation theory is needed. 
Analogous conclusion holds for Y10, for which we explore different modelling choices for the galaxy bias using either the full HEFT or the minimal bias approach in the modelling, shown by the solid pink and blue contours in \autoref{fig:Y10_massive_neutrinos}, respectively. 

We repeat the same analysis using a second mock dataset identical to the first, except that the parameters $b_2$, $\bs$ and $\blap$ are set to twice their values listed in \autoref{tab:fiducials}. The main result, shown in \autoref{fig:Y1_massive_neutrinos} and \autoref{fig:Y10_massive_neutrinos} by the dashed contours, indicates that adding the neutrino mass does not introduce any additional bias in the original cosmological parameters or loss in constraining power.
Nevertheless, when comparing bias modelling choices, we observe that in some mock scenarios the minimal bias can work well and recover the same results as the full HEFT model. However, this behaviour is not robust as for other equally plausible mock scenarios the inferred $M_\nu$ is significantly biased when the minimal bias model is adopted. 
Particularly, we observe a $\sim$$1\sigma$ bias for Y1 and a $\sim$$2\sigma$ bias for Y10 when the mock data with higher values of bias parameters is considered (we see more bias in Y10 because of the higher constraining power and number of lens redshift bins). For Y10, increasing the fiducial bias parameters in this way also changes the FoB for $\Omega_{\rm m}$ and $\sigma_8$ at $\kmax=0.7\,h/$Mpc from $\sim$$0.15\sigma$ to $\sim$$0.85\sigma$ in the minimal bias model. This suggests that the adequacy of the minimal bias description depends sensitively on the galaxy bias values. 

To understand the cause of this shift in $M_\nu$, we repeated the same analysis by modifying only two of $b_2$, $\bs$ and $\blap$ at a time to twice their previous fiducials (these tests are not shown in the plots for clarity). We find that when $b_2$ is the only parameter set to its original fiducial value from \cite{Nicola2024}, the bias in the total neutrino mass is reduced, identifying $b_2$ values in the mock data as the main driver of the shift. Additionally, we note that the original fiducial values in $b_2$ closely follow the $b_1-b_2$ relation for galaxies as in \citet{Zennaro2022}. We leave a closer investigation of the galaxy bias relations for future work.  

\begin{figure}
    \centering
    \includegraphics[width=\linewidth]{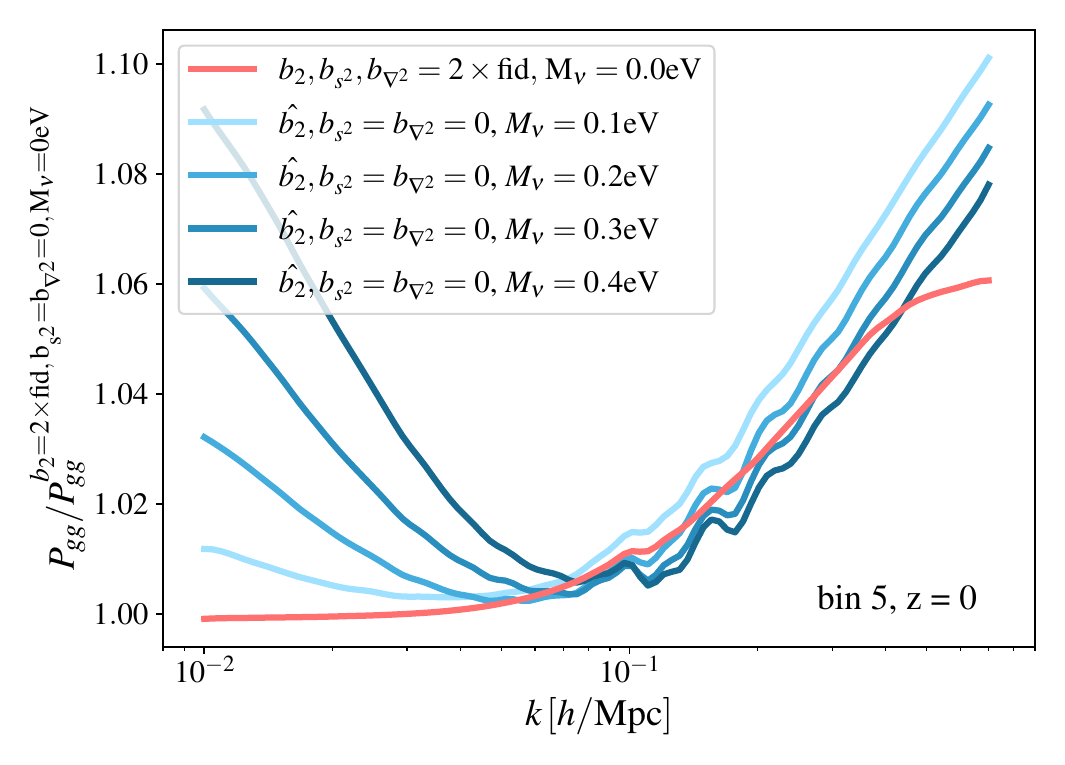}
    \caption{Galaxy-galaxy auto-correlation angular power spectrum at $z=0$ in bin 5 of Y10 showing how bias parameters mimic the effect of massive neutrinos. In pink we show the effect of the bias parameters with no massive neutrinos, where the values of $\bs$ and $\blap$ are double with respect to the fiducial in \autoref{tab:fiducials} to have a more appreciable effect. The blue shaded lines include massive neutrinos and no high order bias parameters ($\bs$ and $\blap$ set to zero). $\hat{b_2}$ indicates the mean value of $b_2$ from the posterior distribution). Both cases are normalised by the angular power spectrum with $b_2$ values double the fiducials, while $\bs$, $\blap$ and $M_\nu$ set to zero.}
    \label{fig:Pgg_neutrinos_bias_Y10}
\end{figure}

Through a series of tests we have found complicated interactions between projection effects, the higher order bias parameters, and the neutrino mass. For example, \autoref{fig:Pgg_neutrinos_bias_Y10} shows the degeneracy between galaxy bias and neutrino mass: we find that the power spectrum with the minimal bias model and massive neutrinos has a similar behaviour to the full HEFT case but with $M_\nu=0\,$eV.
All these degeneracies make it difficult to derive any robust conclusions for general scenarios, and therefore extensive tests should be conducted for any science case before adopting a minimized model. 

However, an important qualitative result is that in each scenario the $M_\nu$ posterior distribution demonstrates a clear peak, distinguishable from zero mass, indicating that LSST Y10 should be able to robustly detect massive neutrino effects even if modelling uncertainties affect the precise value and significance of the inferred value.

Ultimately \autoref{fig:Y10_massive_neutrinos} suggests that LSST should make an unambiguous detection of non-zero neutrino mass, even at its minimal value, regardless of bias modelling and baryons. However, the exact significance of this detection and the precise mass implied will be strongly dependent on these details.

It has been shown that a shear analysis alone is not sufficient for a significant detection of massive neutrinos \cite{Schneider2020b, Tsedrik2024} even in an optimistic Stage IV scenario. The inclusion of semi-nonlinear scales in the galaxy clustering drastically increases the significance of the potential detection. This finding highlights the importance of correctly modelling baryons and nonlinear bias when constraining beyond-$\Lambda$CDM parameters with characteristic nonlinear features.

\section{Conclusions} \label{Sec:Conclusions}
In this work we have studied how baryonic and galaxy bias parameters impact a $3\times2$pt LSST analysis for Y1 and Y10 setups, aiming for a balance between model complexity and scale cuts. Specifically, we reduced the complexity of the baryonic feedback model by fixing all parameters except $\logMc$, and for the bias model we compared three options: linear bias, hybrid effective field theory (HEFT), and a minimal bias model in which the two higher-order HEFT bias parameters are fixed to zero. 
We assumed the same mock data in all cases, generated with the HEFT model for galaxy bias with fiducial values adapted from \cite{Nicola2024}. To assess constraining power and potential biases in the posterior distributions, we computed the figure of merit (FoM) and figure of bias (FoB) for $\Omega_{\rm m}$ and $\sigma_8$. 

For LSST Y1, we found that a linear bias model provides unbiased constraints for $\Omega_{\rm m}$ and $\sigma_8$ (within $0.5\sigma$) up to $k_{\rm max}=0.1\,h/$Mpc, regardless of whether baryonic feedback is included or neglected. However, to extend to smaller scales, a perturbative treatment is required -- either HEFT or the minimal bias model -- which only differ by the number of free bias parameters. We found no significant difference in the FoM of $\Omega_{\rm m}$ and $\sigma_8$ between these two perturbative models (again for either case where baryonic feedback is included or neglected). By contrast, while the qualitative behaviour remained unchanged (HEFT and the minimal bias model demonstrated similar FoM and FoB; with increasing $k_{\rm max}$ FoM was increasing), allowing the baryonic parameter $\logMc$ to vary reduced the FoM by a factor of 2.

Moving from Y1 to Y10 shows an expected improvement by a factor $\sim2$ in the FoM of $\Omega_{\rm m}$ and $\sigma_8$ given by the increased statistical power of the survey.

We demonstrated that while certain combinations of higher-order bias values cannot easily be reproduced by any baryonic feedback scenario, at least within the parameter range allowed by the \bacco ~baryonification model, baryonic suppression on scales relevant for galaxy clustering can generally be mimicked by some choice of higher-order bias parameter. Additionally, we have also tested that neglecting baryonic feedback in the galaxy-galaxy lensing component does not lead to a significant bias in the cosmological parameters but might impact constraints on the baryonic parameters. This conclusion holds both for the full HEFT implementation and the minimal bias model, and for both the Y1 and Y10 configurations.
We note that our analysis is somewhat simplified as we do not consider other systematic effects (\textit{e.g.}, photo-$z$ uncertainties and magnification bias), which might affect neutrino detection \citep{Ribeiro2026}.

Finally, we studied whether LSST will be able to clearly detect the total neutrino mass.
We found that both for LSST Y1 and Y10 a neutrino mass detection is possible going to nonlinear scales, \textit{e.g.}, assuming the minimal bias model and a scale cut $k\gtrsim0.5\,h/$Mpc for Y1.
In LSST Y10 we found that the constraints tighten further (as expected), and the additional free parameter $M_\nu$ does not bias the cosmological parameters.
Analysing different bias prescription both for Y1 and Y10, we found that while a minimal bias model can in some cases reproduce results consistent with the full HEFT framework, this behaviour is not robust and depends sensitively on the assumed galaxy bias values, potentially leading to significant biases the inferred $M_\nu$, and, in some cases, cosmological parameters. These results highlighted the existence of non-trivial degeneracies between neutrino mass, higher-order bias terms, and projection effects, making it difficult to draw general conclusions about simplified modelling approaches. Nevertheless, in all cases considered, the results indicated that LSST should make an unambiguous detection of non-zero neutrino mass even if the precise inferred value is subject to modelling uncertainties.

\section*{Acknowledgements}
This paper has undergone internal review in the LSST Dark Energy Science Collaboration. The internal reviewers were Andrina Nicola and Carlos Garcia-Garcia.

The authors are grateful to Marcos Pellejero Ib{\'a}{\~n}ez for useful insights and discussions.

OT is supported by an Science and Technology Facilities Council (STFC) studentship. MT's and JZ's research is supported by grant ST/Y000986/1. AP is a UK Research and Innovation Future Leaders Fellow [grant MR/X005399/1]. AP's research is supported by a UK Research and Innovation Future Leaders Fellowship [grant MR/X005399/1]. NS is supported in part by the OpenUniverse effort, funded by NASA under JPL Contract Task 70-711320, ``Maximizing Science Exploitation of Simulated Cosmological Survey Data Across Surveys''. For the purpose of open access, the author has applied a Creative Commons Attribution (CC BY) license to any Author Accepted Manuscript version arising from this submission. 
CG is funded by the MICINN project PID2022-141079NB-C32. IFAE is partially funded by the CERCA program of the Generalitat de Catalunya.
The DESC acknowledges ongoing support from the Institut National de Physique Nucl\'eaire et de Physique des Particules in France; the Science \& Technology Facilities Council in the United Kingdom; and the Department of Energy and the LSST Discovery Alliance in the United States.  DESC uses resources of the IN2P3 Computing Center (CC-IN2P3--Lyon/Villeurbanne - France) funded by the Centre National de la Recherche Scientifique; the National Energy Research Scientific Computing Center, a DOE Office of Science User Facility supported by the Office of Science of the U.S.\ Department of Energy under Contract No.\ DE-AC02-05CH11231; STFC DiRAC HPC Facilities, funded by UK BEIS National E-infrastructure capital grants; and the UK particle physics grid, supported by the GridPP Collaboration. This work was performed in part under DOE Contract DE-AC02-76SF00515.

{\bf Author contributions.} OT co-wrote the analysis pipeline and wrote the post-processing code, performed analysis, and wrote the paper. MT coded the analysis pipeline, contributed to discussion and paper writing. JZ and AP initialized
project idea, provided comments and supervision throughout
project, and reviewed the manuscript. NS provided synthetic $n(z)$ and analysis-specific fiducial values of nonlinear galaxy bias parameters, and reviewed the manuscript. CG is a DESC builder and supervised the project as group convenor.

\section*{Data Availability}
Links and references with sources of the data and analysis pipeline are provided in the main text. 
The MGL pipeline used in the entire analysis, and its comparison to CCL (\href{https://github.com/LSSTDESC/CCL/}{https://github.com/LSSTDESC/CCL/}) pipeline, is available at \href{https://github.com/MariaTsedrik/MGlensing}{https://github.com/MariaTsedrik/MGlensing}. The small-scales systematics are modelled using the BACCO emulator \href{https://bitbucket.org/rangulo/baccoemu/src/master/}{https://bitbucket.org/rangulo/baccoemu}. The SRD forecasting data are available at \href{https://github.com/LSSTDESC/forecasting}{https://github.com/LSSTDESC/forecasting}. The best fit values in \autoref{fig:avg_std_y1_sqrtS} are calculated using \texttt{iminuit}, \href{https://github.com/scikit-hep/iminuit}{https://github.com/scikit-hep/iminuit}

\appendix

\section{A. Comparison with CCL}\label{Appendix:ccl}
\begin{figure}
    \centering
    \includegraphics[width=\textwidth]{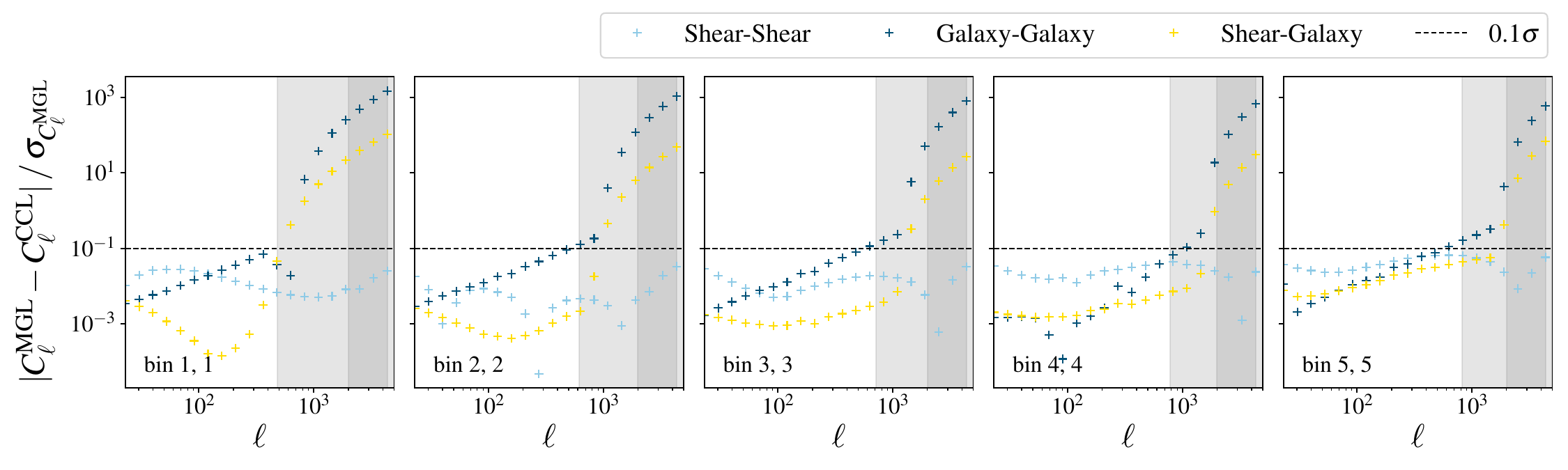} 
    \caption{The absolute value of the difference between the $C_\ell$ obtained with \texttt{CCL} and \texttt{MGL} weighted by the error $\sigma_{C_\ell}$ for the shear-shear (light blue), galaxy-galaxy (blue) and galaxy-shear (yellow) angular power power spectra. The dashed horizontal line shows the $10\%$ of $1\sigma$ error, the target numerical accuracy from \cite{Chisari2019}. The light grey shaded area corresponds to the galaxy-galaxy and galaxy-shear scale cuts $\kmax=0.7\, h/$Mpc, while the dark grey area corresponds to the shear-shear scale cuts at $\ell_{\rm max}=2000$.}
    \label{fig:ccl_mgl_cells}
\end{figure}

\begin{figure}
    \centering
    \includegraphics[width=0.45\textwidth]{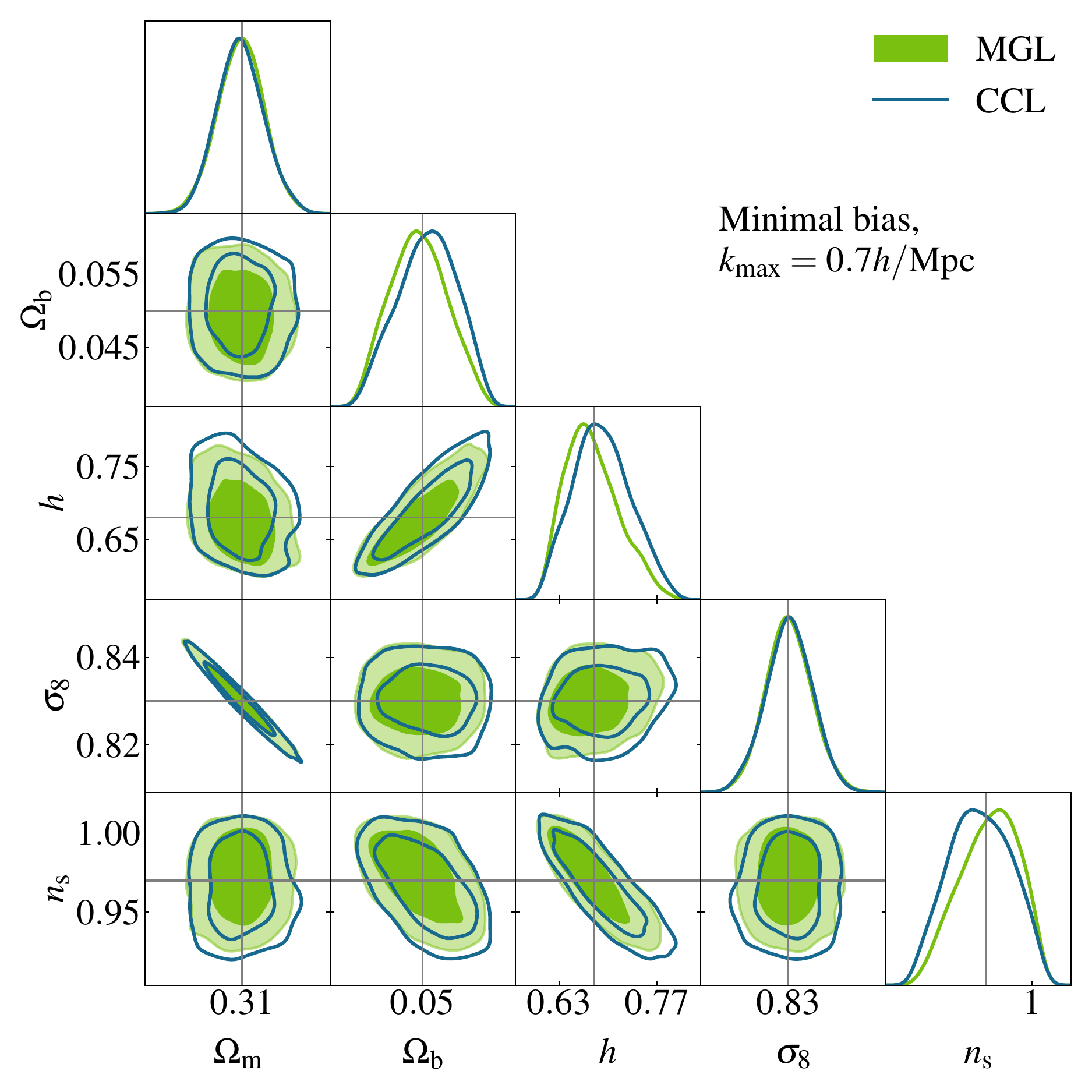}
    \caption{The 68\% and 95\% 2D confidence regions and posteriors for some of the parameters in the analysis at $\kmax=0.7\,h/$Mpc for LSST Y1. The labels \texttt{CCL} (blue contours) and \texttt{MGL} (green contours) indicate the pipeline used to compute the $C_\ell$s both in the analysis and in the data vector.}
    \label{fig:ccl_mgl_posterior}
\end{figure}

We show here the comparison between $C_\ell$s computed in the \texttt{MGL} pipeline with those from the DESC Core Cosmology Library (\texttt{CCL})\footnote{\href{https://github.com/LSSTDESC/CCL}{https://github.com/LSSTDESC/CCL}} \citep{Chisari2019} -- both with the \bacco ~emulator for the nonlinear matter-matter and the HEFT galaxy-galaxy and galaxy-matter power spectra.
\autoref{fig:ccl_mgl_cells} shows that the differences between the $C_\ell$s obtained from \texttt{CCL} and \texttt{MGL} are always well within  $10\%$ of the $1\sigma$ error bars within the scale cuts considered in this work (LSST Y1 in the plot, but same consideration holds for Y10). The small difference in the galaxy-galaxy spectrum at small scales is due to the different extrapolation in $k$ used in the two methods.  
We also found that the total $\Delta\chi^2$ is 0.15 for Y1 and 1.35 for Y10, indicating that the differences between the two implementations are negligible for the purposes of this analysis.
In \autoref{fig:ccl_mgl_posterior} we compare the parameter posterior distributions obtained using the same implementation in the data and in the analysis. 
The weakly constrained parameters ($\Omega_{\rm b},\, h$, and $n_{\rm s}$) show the largest difference, for which the shift between the two posterior is $\sim20\%$ of the $1\sigma$ uncertainty in the most extreme case. This is likely due to the different implementation of extrapolation in high $k$ and high $z$ outside the emulated \bacco ~range. 
For all other parameters, the agreement is better, confirming that the differences between the \texttt{CCL} and \texttt{MGL} implementations have a negligible impact on our conclusions.

\section{B. Quantitative bias-baryon degeneracy}\label{Appendix:quantify_rainbow}
To quantify the possible similarity between higher-order bias and baryonic effects illustrated in the \autoref{fig:rainbow_plot} we run the minimiser {\tt iminuit}\footnote{\url{https://iminuit.readthedocs.io/en/stable/index.html}}
\citep{iminuit} in the following set-up: we generate a $C_\ell^{\rm GG}$ mock LSST-Y1 data vector using the Eulerian linear bias and nonlinear matter power spectrum with baryonic suppression. We fit it with a model computed at the same cosmology, but with the HEFT model for the corresponding Lagrangian linear bias and varying $b_2, b_{s^2}, b_{\nabla^2}$, but with no baryon effects (leading to 15 varied parameters). 

\begin{figure}
    \centering
    \includegraphics[width=\textwidth]{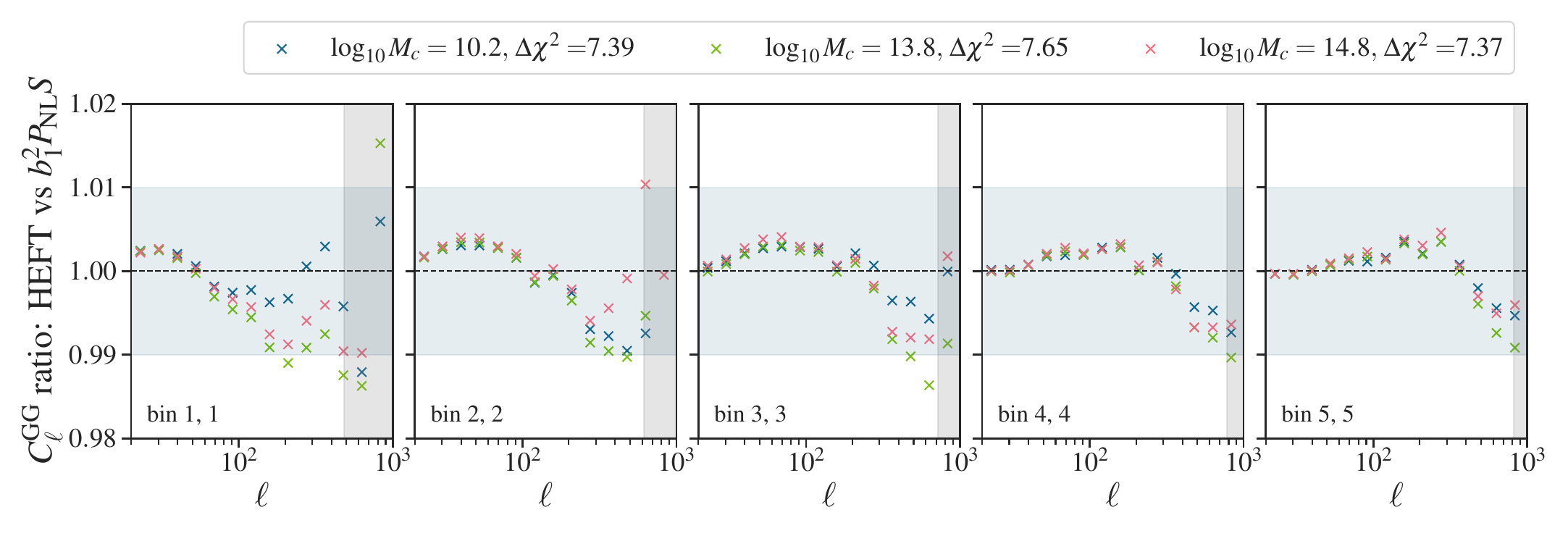} 
    \caption{The ratios between galaxy clustering $C_\ell$ computed in the HEFT model with respect to the naive modelling via $b_1^2P_{\rm NL}S$ with weak (blue), medium (green) and strong (pink) baryonic suppression. The light grey shaded area corresponds to the galaxy-galaxy scale cuts $\kmax=0.7\, h/$Mpc, while the blue shaded area corresponds to 1\% accuracy.}
    \label{fig:gg_cells_baryons_heft}
\end{figure}

Best-fit results for different strengths of baryonic suppression in the mock data are shown in \autoref{fig:gg_cells_baryons_heft} with the corresponding goodness-of-fit values shown above. We show only the auto-correlated bins, while the fit and $\Delta \chi^2$ values are computed for all 15 redshift-bins, cross-correlations included, \textit{i.e.}, the length of the data-vector is 178 for the galaxy-galaxy scale cuts $\kmax=0.7\, h/$Mpc. From this we conclude that it is possible to find a combination of higher-order galaxy bias parameters, which can mimic baryonic suppression with 1\% accuracy.   

\section{C. Comparison with total covariance}\label{Appendix:tot_cov}
\begin{figure}
    \centering
    \includegraphics[width=0.45\textwidth]{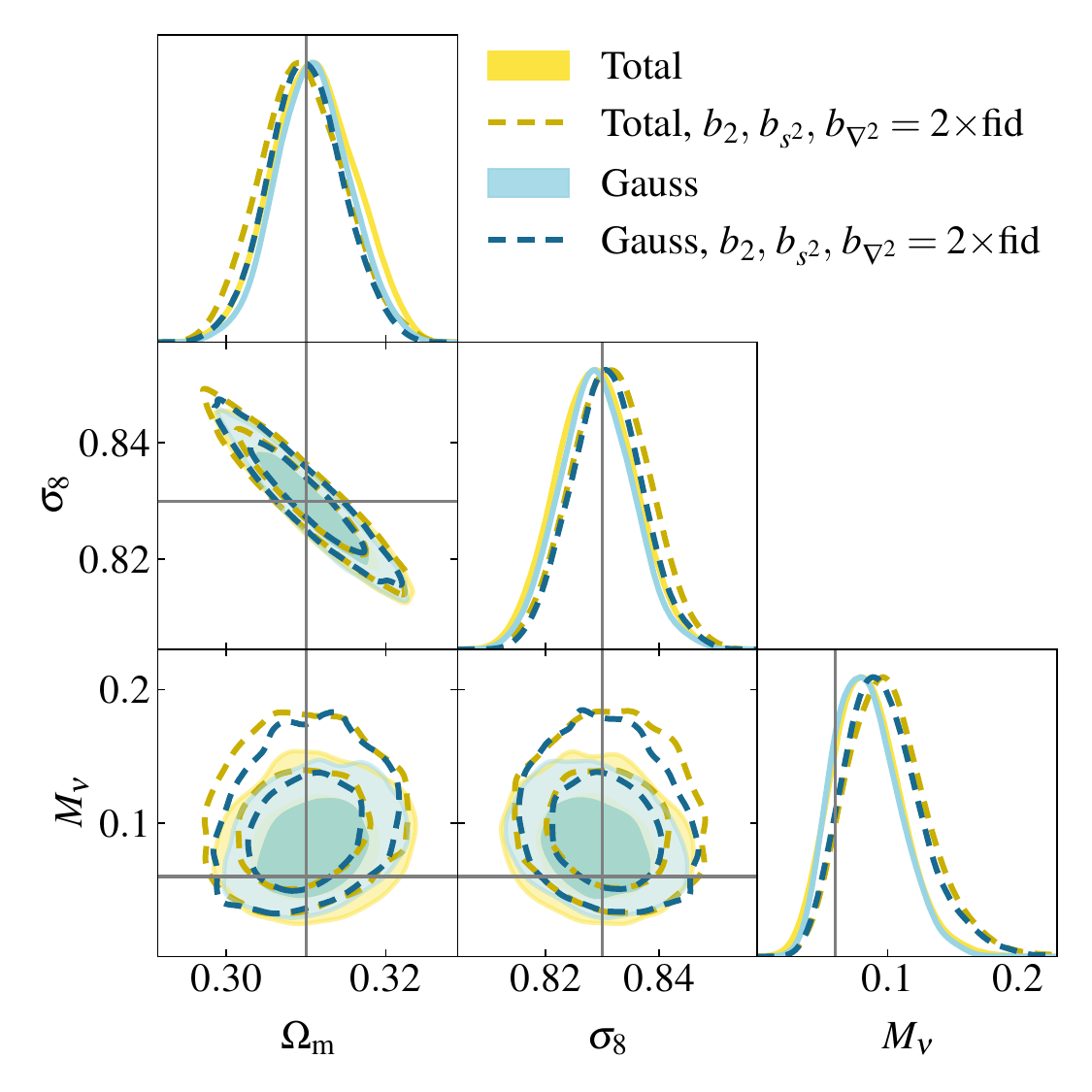}
    \caption{The 68\% and 95\% 2D confidence regions and posteriors for some of the parameters in the analysis at $\kmax=0.7\,h/$Mpc for LSST Y1. We assumed HEFT in the mock data and the minimal bias model in the analysis, both with baryonic feedback and massive neutrinos (free $\logMc$ and $M_\nu$). The solid contours consider the fiducial values as in \autoref{tab:fiducials}, while the dashed blue contour consider the fiducial values in the mock data of $b_2$, $\bs$ and $\blap$ to be twice as those. The blue and yellow contours indicate the results obtained with a Gaussian and a total covariance, respectively.}
    \label{fig:total_cov_Mnu}
\end{figure}

As described in \autoref{Subsec:mgl}, the pipeline used in this work includes a Gaussian covariance. However, given their high precision measurements, Stage IV surveys will be sensitive to additional non-Gaussian terms in the covariance matrix \citep{Barreira2018, Sciotti2024}. To test the validity of our conclusions, we repeat our analysis using a non-Gaussian covariance and compare the corresponding results with the ones presented in the main text of this work.
Specifically, we use the publicly available code \texttt{Spaceborne} (SB)
\footnote{\href{https://github.com/davidesciotti/Spaceborne}{https://github.com/davidesciotti/Spaceborne}},
to compute the total covariance (\textit{i.e.}, with Gaussian super-sample covariance and connected non-Gaussian terms). We find that all results and conclusions presented in this paper are largely unaffected by the inclusion of non-Gaussian terms. For instance, in the scale-cut runs from \autoref{sec:scalecuts}, the main difference is found in the $\Omega_{\rm m}-\sigma_8$ FoM in the HEFT scenario, which is approximately 10\% lower when using the total covariance. Additionally, we test whether our conclusions on the detection of massive neutrinos and on the impact of the fiducial galaxy-bias values might be sensitive to the broadening of the posterior induced by the additional covariance terms. \autoref{fig:total_cov_Mnu} shows a comparison between the total neutrino mass posterior obtained using the Gaussian and total covariances. Evidently, the conclusions on massive neutrinos from \autoref{sec:neutrinos} are not affected by the use of the total covariance.

\section{D. Additional baryonic parameters}\label{Appendix:baryons}
As a robustness check, we now compare the results presented in \autoref{Subsubs:simplifyingbias} with a more conservative scenario where the full set of \texttt{BACCO}'s baryonic parameters ($\eta$,\, $\theta_{\rm inn},\, M_{\rm inn},\, \theta_{\rm out}, M_{\rm c},\, \beta, \,M_{\rm 1,z0,cen}$) is allowed to vary, instead of just $M_{\rm c}$, as in our fiducial analysis. We tested both the full HEFT and minimal bias scenarios.

We find that while the FoM has similar overall dependence on $\kmax$ in the four cases, its values are around a factor of two lower when all the parameters are freed. On the other hand, the FoB values are systematically slightly larger in the extended scenario, with a maximum of FoB $\sim0.84\sigma$ at $\kmax=0.3\,h/$Mpc for the minimal bias model case.

\begin{figure}
    \centering
    \includegraphics[width=0.45\linewidth]{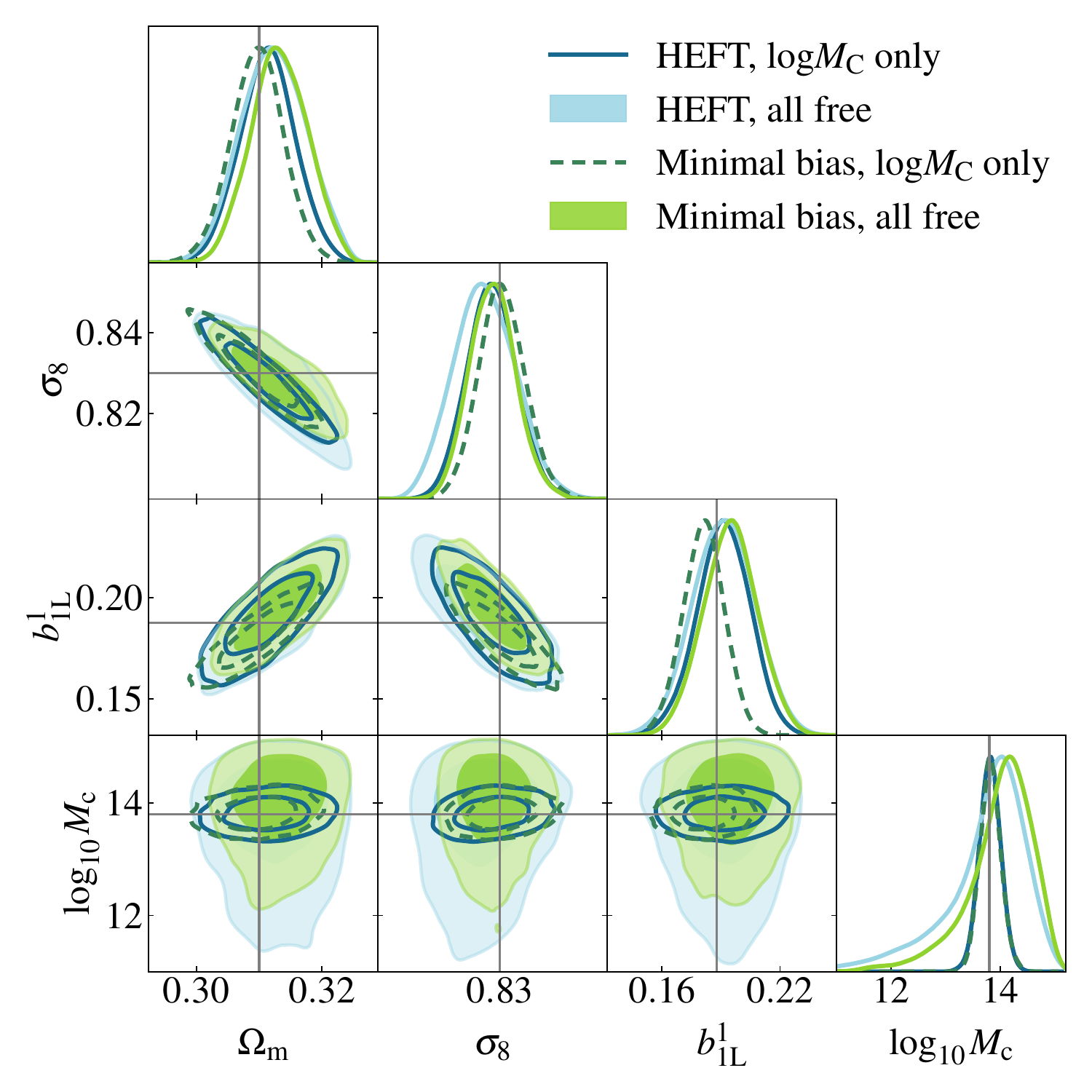} 
    \caption{The 68\% and 95\% 2D confidence regions and posteriors for some of the free parameters in the analysis at $\kmax=0.7\,h/$Mpc, considering HEFT and minimal bias as bias model. Light blue and light green results indicate the analysis used in this work, which considers $\logMc$ as only free baryonic parameter, while in dark blue and dark green indicate the cases where all baryonic parameters or only three of them are allowed to vary.}
    \label{fig:all-baryons-free}
\end{figure}

\autoref{fig:all-baryons-free} shows the posterior distribution for some of the sampled parameters in the analysis, while \autoref{tab:fom_fob_baryons} reports the FoM and FoB for $\Omega_{\rm m}$ and $\sigma_8$ in all these scenarios for $\kmax=0.3$ and $0.7\,h/$Mpc. 
Allowing additional baryonic parameters to vary affects both cosmological and nuisance-parameter constraints. In particular, the constraints on $\Omega_{\rm m}$ and $\sigma_8$ become broader and less precise, especially when assuming the minimal bias model. These broadening and shifts arise from additional degeneracy directions and projection effects introduced by the wider baryonic posterior distribution, which propagates into the the cosmological and nuisance posteriors. In particular, as seen in \autoref{Subsubs:simplifyingbias}, six out of the seven baryonic parameters cannot be constrained by $3\times2$pt data alone and require the use of external data.
This indicates that the quantitative conclusions (\textit{i.e.}, FoM and FoB values, recommendations on scale-cuts) from our reference analysis, in which only $\logMc$ is varied, could change when the baryonic parameter space is extended. This impact is even more significant on bias parameters, reducing our ability to constrain these nuisance parameters with precision.

Overall, this robustness test indicates that a completely free baryon model has major effects on our ability to constrain any parameters, but also that a single parameter model in this form may not be sufficient for Stage IV. Cautious treatments of upcoming data should carefully develop and select priors for the full \texttt{BACCO} parameter space, especially if adopting minimal bias models.

\begin{table}
    \centering
    \begin{tabular}{cccccc}
      & $\kmax\, [h/$Mpc] & FoM ($\logMc$) & FoM (all free) & FoB ($\logMc$) & FoB (all free) \\
    \hline
     HEFT & 0.3 & $6.90\times 10^4$ & $3.62\times 10^4$ & 0.12 &  0.56 \\
     HEFT & 0.7 & $9.45\times 10^4$ & $4.57\times 10^4$ & 0.32 & 0.49 \\
     minimal bias & 0.3 & $7.00\times 10^4$ & $3.59\times 10^4$ & 0.08 & 0.61 \\
     minimal bias & 0.7 & $1.02\times 10^5$ & $6.00\times 10^4$ & 0.08 & 0.84 \\
     \hline
    \end{tabular}
    \caption{$\Omega_{\rm m}$ and $\sigma_8$ figure of merit and figure of bias with $\kmax=0.3$ and $0.7\, h/$Mpc for cases where either all baryonic parameter or $\logMc$ only are allowed to vary.}
    \label{tab:fom_fob_baryons}
\end{table}

\section{E. Additional figures and tables}
\label{appendix-extra-figs-tables}
We collect here the full posterior distributions for a few meaningful runs described in the text, as well as the fiducial values and priors for all the parameters used in the analysis.

\begin{figure}
    \centering
    \includegraphics[width=\textwidth]{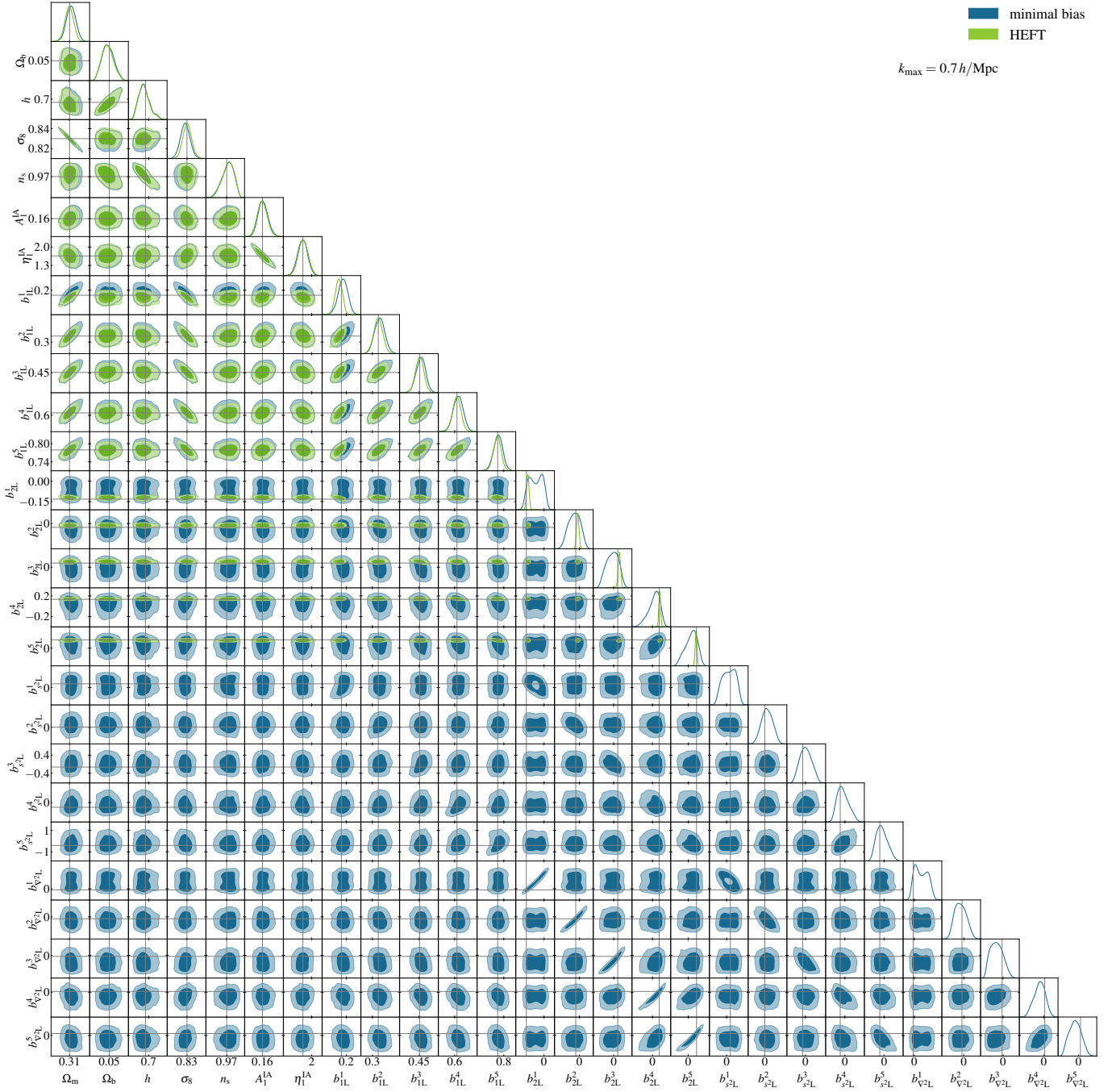}
    \caption{The 68\% and 95\% 2D confidence regions and posteriors for all free parameters in the analysis at $\kmax=0.7\,h/$Mpc. Both runs consider the same mock data set with \textit{Zennaro24} and HEFT bias model, but different analysis model: HEFT (blue) and minimal bias (green).}
    \label{fig:corner-heft-fix-kmax0.7}
\end{figure}

\begin{figure}
    \centering
    \includegraphics[width=0.45\textwidth]{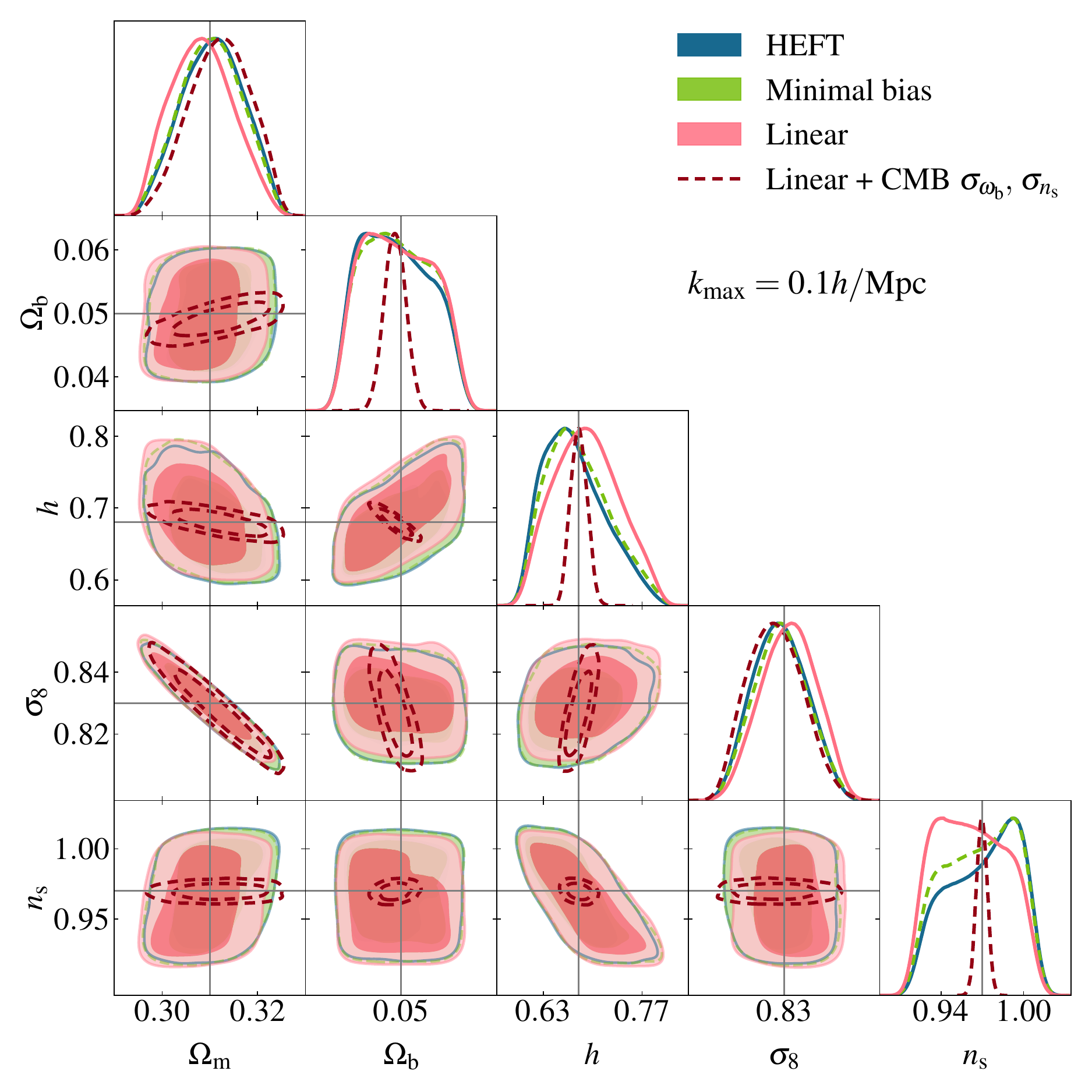} \quad
    \includegraphics[width=0.45\textwidth]{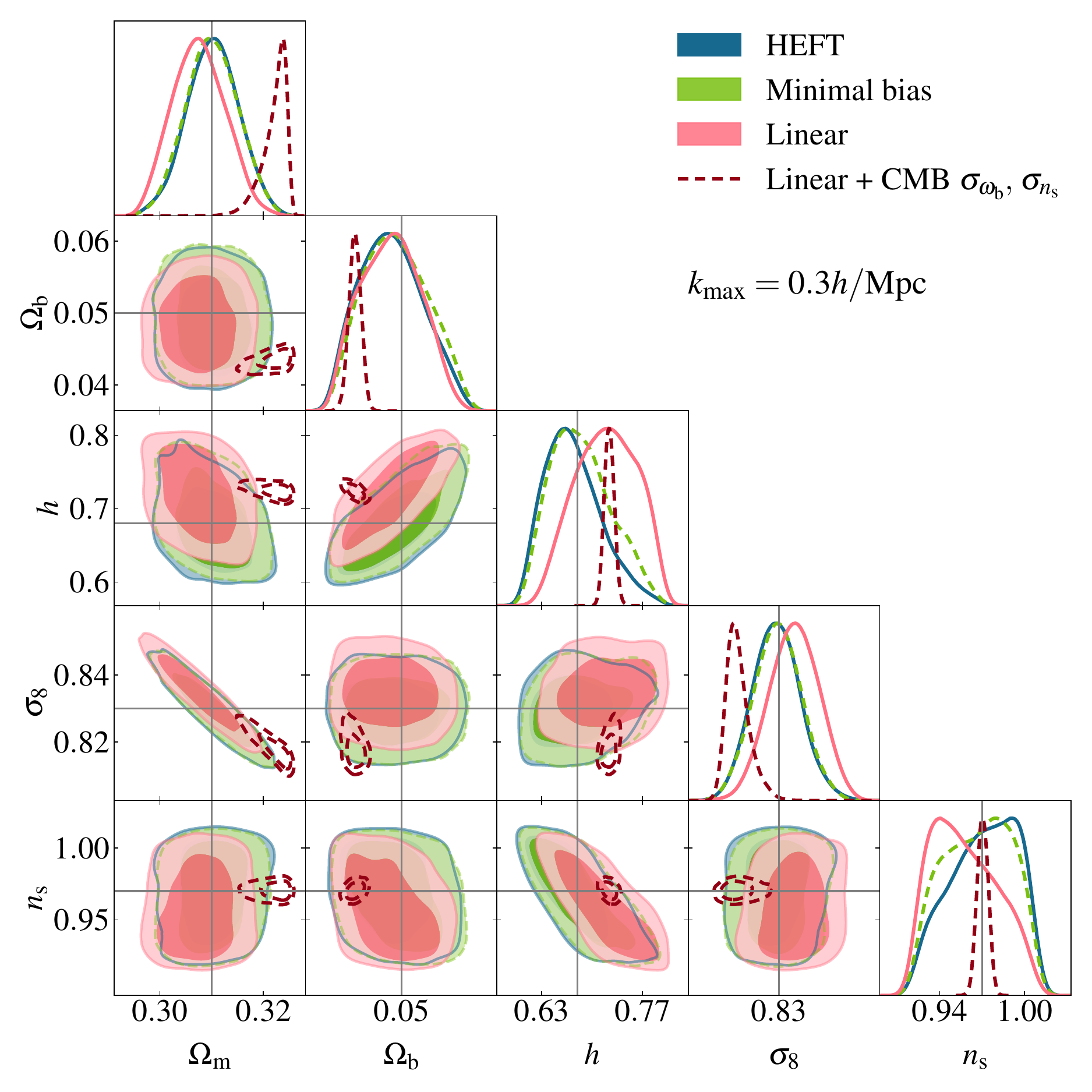}
    \caption{The 68\% and 95\% 2D confidence regions and posteriors for cosmological and IA parameters in the analysis at $\kmax=0.1\,h/$Mpc (left) and $\kmax=0.3\,h/$Mpc (right). 
    All runs consider the same mock data set with HEFT bias model, which are then analysed with three different models: HEFT (blue), minimal bias model (green) and linear bias (pink and dashed red contours). For the case represented by the dashed red contour we assumed CMB-like prior on $\Omega_{\rm b}h^2=\omega_{\rm b}$ and $n_{\rm s}$.}
    \label{fig:corner-b1-heft-fix-kmax0.3}
\end{figure}

\begin{table}
    \centering
    \begin{tabular}{ccc}
    Parameter & Fiducial & Prior \\
    \hline
    $\Omega_{\rm m}$ & 0.31 & [0.24, 0.39] \\
    $\sigma_8$       & 0.83 & [0.73, 0.9] \\
    $h$              & 0.68 & [0.6, 0.8] \\
    $\Omega_{\rm b}$ & 0.05 & [0.04, 0.06] \\
    $n_{\rm s}$      & 0.97 & [0.92, 1.01] \\
    $M_\nu$          & 0.06 & [0, 0.4] \\
    $A^{\rm IA}_1$    & 0.16 & [-1, 5] \\
    $\eta^{\rm IA}_1$ & 1.66 & [-5, 5] \\
    \hline
    $\logMc$ & 13.8 & [9, 15] \\
    log$_{10}\eta$ & -0.3 &  [-0.69, 0.69] \\
    log$_{10}\beta$ & -0.22  &  [-1, 0.69] \\
    log$_{10}M_{1,z0,cen}$ & 10.5 &  [9, 13] \\
    log$_{10}\theta_{\rm out}$ & 0.25 &  [0, 0.47] \\
    log$_{10}\theta_{\rm in}$ & -0.86 &  [-2, -0.52] \\
    log$_{10}M_{\rm inn}$ & 12.4 &  [9, 13.5] \\ 
    \hline 
    & & \\
    Parameter (Y1) & Fiducial & Prior \\
    \hline
    $b_{1,1}$ & 0.187684 & [0, 3] \\
    $b_{1,2}$ & 0.312375 & [0, 3] \\
    $b_{1,3}$ & 0.45121  & [0, 3] \\
    $b_{1,4}$ & 0.60626  & [0, 3] \\
    $b_{1,5}$ & 0.779385 & [0, 3] \\
    $b_{2,1}$ & -0.1322075 & [-2, 2] \\
    $b_{2,2}$ & -0.0345375 & [-2, 2] \\
    $b_{2,3}$ & 0.0579305  & [-2, 2] \\
    $b_{2,4}$ & 0.133662   & [-2, 2] \\
    $b_{2,5}$ & 0.232387   & [-2, 2] \\
    $b_{\rm s^2,1}$ & 0.1127485  & [-2, 2] \\
    $b_{\rm s^2,2}$ & 0.0036835  & [-2, 2] \\
    $b_{\rm s^2,3}$ & -0.1261985 & [-2, 2] \\
    $b_{\rm s^2,4}$ & -0.285537  & [-2, 2] \\
    $b_{\rm s^2,5}$ & -0.406101  & [-2, 2] \\
    $b_{\nabla^2,1}$ & -0.100126  & [-2, 2] \\
    $b_{\nabla^2,2}$ & -0.1563475 & [-2, 2] \\
    $b_{\nabla^2,3}$ & -0.115281  & [-2, 2] \\
    $b_{\nabla^2,4}$ & 0.1059295  & [-2, 2] \\
    $b_{\nabla^2,5}$ & 0.361450   & [-2, 2] \\
    \hline    
    \end{tabular}
    \quad 
    \begin{tabular}{ccc}
    Parameter (Y10)  & Fiducial & Prior \\
    \hline
    $b_{1,1}$ & 0.187684 & [0, 3] \\
    $b_{1,2}$ & 0.213704 & [0, 3] \\
    $b_{1,3}$ & 0.279298 & [0, 3] \\
    $b_{1,4}$ & 0.345371 & [0, 3] \\
    $b_{1,5}$ & 0.414455 & [0, 3] \\
    $b_{1,6}$ & 0.492622 & [0, 3] \\
    $b_{1,7}$ & 0.570881 & [0, 3] \\
    $b_{1,8}$ & 0.650379  & [0, 3] \\
    $b_{1,9}$ & 0.740059  & [0, 3] \\
    $b_{1,10}$ & 0.829755 & [0, 3] \\
    $b_{2,1}$  & -0.1322075 & [-2,2] \\
    $b_{2,2}$  & -0.111826  & [-2,2] \\
    $b_{2,3}$  & -0.060447  & [-2,2] \\
    $b_{2,4}$  & -0.008693  & [-2,2] \\
    $b_{2,5}$  & 0.0399785  & [-2,2] \\
    $b_{2,6}$  & 0.078157   & [-2,2] \\
    $b_{2,7}$  & 0.1163815  & [-2,2] \\
    $b_{2,8}$  & 0.15617    & [-2,2] \\
    $b_{2,9}$  & 0.209154   & [-2,2] \\
    $b_{2,10}$ & 0.262147   & [-2,2] \\
    $b_{\rm s^2,1}$  & 0.1127485  & [-2,2] \\
    $b_{\rm s^2,2}$  & 0.089989   & [-2,2] \\
    $b_{\rm s^2,3}$  & 0.0326155  & [-2,2] \\
    $b_{\rm s^2,4}$  & -0.0251765 & [-2,2] \\
    $b_{\rm s^2,5}$  & -0.088427  & [-2,2] \\
    $b_{\rm s^2,6}$  & -0.1687555 & [-2,2] \\
    $b_{\rm s^2,7}$  & -0.2491795 & [-2,2] \\
    $b_{\rm s^2,8}$  & -0.326995  & [-2,2] \\
    $b_{\rm s^2,9}$  & -0.3819875 & [-2,2] \\
    $b_{\rm s^2,10}$ & -0.436989  & [-2,2] \\
    $b_{\nabla^2,1}$  & -0.100126 & [-2,2] \\
    $b_{\nabla^2,2}$  & -0.111858 & [-2,2] \\
    $b_{\nabla^2,3}$  & -0.141433 & [-2,2] \\
    $b_{\nabla^2,4}$  & -0.171224 & [-2,2] \\
    $b_{\nabla^2,5}$  & -0.167719 & [-2,2] \\
    $b_{\nabla^2,6}$  & -0.056198 & [-2,2] \\
    $b_{\nabla^2,7}$  & 0.0554545 & [-2,2] \\
    $b_{\nabla^2,8}$  & 0.1694515 & [-2,2] \\
    $b_{\nabla^2,9}$  & 0.3029225 & [-2,2] \\
    $b_{\nabla^2,10}$ & 0.436416  & [-2,2] \\
    \hline
    \end{tabular}
    \caption{Fiducials and uniform prior ranges of the parameters used in this work. For fiducial values we used \textit{Planck}-like cosmological parameters \citep{PlanckCollaboration2020}} and bias parameters as in \citetalias{Mandelbaum2018}. We chose baryonic feedback parameters to have a moderate suppression. Note that we vary $A_1^{\rm IA}=A_1/(1+z_{\rm piv})^{\eta_1}$ and $\eta_1^{\rm IA}=\eta_1$ from \autoref{eq:IA}.  The prior range for cosmological and $\logMc$ parameters follows the \bacco ~emulator setup.
    \label{tab:fiducials}
\end{table}

\bibliographystyle{mnras_2author}
\bibliography{references.bib}

\end{document}